\documentclass[twocolumn,aps,pre,amsmath,showpacs]{revtex4-1}
\usepackage{graphicx,dcolumn,bm,amssymb,amsmath,ulem,indentfirst,amsthm}
\usepackage[toc,page]{appendix}
\usepackage{color}

\theoremstyle{plain}
\def\be{\begin{equation}}
\def\ee{\end{equation}}

\begin{document}

\author{Alessio Zaccone$^{1,2,3}$}
\affiliation{${}^1$Department of Physics ''A. Pontremoli", University of Milan, via Celoria 16, 20133 Milano, Italy.}
\affiliation{${}^2$Statistical Physics Group, Department of Chemical
Engineering and Biotechnology, University of Cambridge, New Museums Site, CB2
3RA Cambridge, U.K.}
\affiliation{${}^3$Cavendish Laboratory, University of Cambridge, JJ Thomson
Avenue, CB30HE Cambridge,
U.K.}
\begin{abstract}

The relaxation dynamics and the vibrational spectra of amorphous solids, such as metal alloys, have been intensely investigated as well separated topics in the past. The aim of this review is to summarize recent results in both these areas in an attempt to establish, or unveil, deeper connections between the two phenomena of relaxation and vibration. Theoretical progress in the area of slow relaxation dynamics of liquid and glassy systems and in the area of vibrational spectra of glasses and liquids is reviewed.
After laying down a generic modelling framework to connect vibration and relaxation, the physics of metal alloys is considered where the emergence of power-law exponents has been identified both in the vibrational density of states (VDOS) as well as in density correlations. Also, theoretical frameworks which connect the VDOS to the relaxation behaviour and mechanical viscoelastic response in metallic glasses are reviewed. The same generic interpretative framework is then applied to the case of molecular glass formers where the emergence of stretched-exponential relaxation in dielectric relaxation can be put in quantitative relation with the VDOS by means of memory-function approaches. Further connections between relaxation and vibration are provided by the study of phonon linewidths in liquids and glasses, where a natural starting point is given by hydrodynamic theories.
Finally, an agenda of outstanding issues including the appearance of compressed exponential relaxation in the intermediate scattering function of experimental and simulated systems (metal alloys, colloidal gels, jammed packings) is presented in light of available (or yet to be developed) mathematical models, and compared to non-exponential behaviour measured with macroscopic means such as mechanical spectroscopy/rheology.

\end{abstract}

\title{Relaxation and vibrational properties in metal alloys and other disordered systems}
\maketitle

\section{Introduction}
\subsection{Relaxation}
In high-temperature liquids, where ''high-temperature" here refers to temperatures significantly higher than the freezing transition and the glass transition, the intermediate scattering function $F(q,t)$, defined as the Fourier transform of density fluctuations in the fluid, exhibits a simple exponential decay $\sim \exp (-t/\tau)$, which reflects an underlying simple diffusive dynamics of the building blocks (atoms, molecules). 
Upon lowering the temperature close to the freezing (crystallization) or the glass transition, liquids become very viscous, with a spectacular increase of viscosity in the supercooled regime by up to 13 orders of magnitude. In the supercooled regime, the atomic motions become highly correlated and each atom becomes effectively ''caged" on a time-scale which becomes larger and larger. Mean-field models such as the Mode-Coupling Theory predict that this timescale diverges at a critical temperature, the Mode-Coupling temperature $T_{c}$~\cite{Goetze_1992}, however there is no evidence of diverging quantities in experiments.

In the supercooled state, the decay of density correlations on long time scales is collective and is no longer described by a simple exponential, but rather by a stretched-exponential function,
$\sim \exp (-t/\tau)^{\beta}$, with a stretching exponent $\beta <1$ which can vary depending on temperature, system composition and many other factors. This process is known as $\alpha$-relaxation.  

Pioneering ideas by J. Frenkel in the 1930s~\cite{Frenkel1,Frenkel2} established a link between the $\alpha$ relaxation time and Maxwell viscoelasticity, by identifying the $\alpha$ relaxation time $\tau$ with the time necessary for an atom to escape the nearest-neighbour cage, $\tau = \tau_{0} \exp(U/k_{B}T)$. Here $\tau_0$ is an attempt frequency, while $U$ is an activation energy barrier which separates two minima (basins) in the energy landscape. Clearly, in the supercooled regime this escape-from-the-cage process is highly cooperative and collective because the dynamics of many atoms is nonlinearly coupled through a feedback mechanism that is quantitatively captured by the framework of Mode-Coupling Theory~\cite{Goetze}.

A second decay of $F(q,t)$ may happen on a shorter time-scale in certain systems, which is known as $\beta$ or secondary relaxation. While $\alpha$ relaxation represents a global or collective process, the $\beta$ relaxation is localized and has been variously associated with either molecular degrees of freedom or with string-like motion of a small number of atoms/molecules.

More recently, non-exponential relaxation $\sim \exp (-t/\tau)^{\beta}$ with $\beta > 1$ (or compressed exponential relaxation) has been observed in several systems including, notably, metal alloys both in experiments~\cite{Ruta} and in simulations~\cite{Kob2018}. This form of the decay of $F(q,t)$ can be accompanied by ballistic dynamical signatures such as a relaxation time that scales as $\tau \sim q^{-1}$, in contrast with the scaling $\tau \sim q^{-2}$ typical of diffusion. 

In general, as shown by molecular simulations~\cite{Sciortino}, the contributions from the inherent structures (i.e.  the potential energy basins) to the equilibrium free energy of supercooled liquids completely decouples from the vibrational contribution. Hence, the relation between vibration and relaxation requires conceptual tools from nonequilibrium statistical mechanics. In the following, various phenomenological facts and interpretative/modelling frameworks are reviewed, starting from the vibrational properties.

\subsection{Vibrational spectra  of liquids and glassy solids}
On the other hand, much of previous research in condensed matter structure and dynamics has focused on the vibrational properties of amorphous systems, including amorphous materials, viscous supercooled liquids and normal liquids. The key quantity here is the vibrational density of states (VDOS), which is defined as the normalized distribution of eigenfrequencies of a material. For a solid, the VDOS is directly related to the distribution of eigenvalues of the Hessian (dynamical)  matrix, and it is generally semi-positive defined being the material rigid. For liquids, which are not rigid but flow under an applied stress, negative eigenvalues occur, as is well known, which result in purely imaginary frequencies in the VDOS, also known as instantaneous normal modes~\cite{Stratt}. Although these eigenfrequencies cannot be accessed experimentally, they are important as they represent non-propagating exponentially-decaying, hence relaxing modes. These imaginary modes physically describe the local relaxation of unstable saddle points which are continuously generated by thermal fluctuations in the liquid.

The VDOS of solids can be measured experimentally using inelastic scattering techniques, such as X-ray and neutron scattering, where the dynamic structure factor $S(q,\omega)$ is directly measured. Upon summing over all wavevectors $q$, the VDOS is obtained. In particular, for metallic glasses and metal alloys inelastic neutron scattering and inelastic X-ray scattering are the most used techniques thanks to the favourable contrast for, respectively, isotopic and electronic scattering length. 
For oxide and for organic glasses formed by either small molecules or polymers, Raman and Brillouin scattering techniques have also been used extensively. 
 
The accessible part of the vibrational spectrum of liquids measured in inelastic scattering experiments presents a symmetric structure for positive and negative energy transfer, with a diffusive Rayleigh peak (quasi-elastic contribution) due to spontaneous thermal fluctuations, and then two symmetric Brillouin doublets related to acoustic contributions. The structure of the liquid spectrum has been derived using hydrodynamics (conservation equations) by Landau and Placzek ~\cite{Placzek} and later on in the frame of generalized or molecular hydrodynamics~\cite{Yip}.  

In glassy solids, the low-frequency part of the VDOS behaves in agreement with the Debye law $\sim \omega^{2}$ from $\omega=0$ up to a crossover frequency at which phonons no longer propagate ballistically but start to feel the mesoscopic disordered structure which causes phonon scattering. The combination of disorder-induced and anharmonic scattering leads through a Ioffe-Regel  crossover into a regime where vibrational modes propagate in a diffusive-like manner (diffusons). At this crossover, the VDOS normalized by $\omega^{2}$ features a peak which is known as the boson peak. Historically, it was shown early on in Raman scattering experiments that this peak seems to obey the same frequency and temperature dependence as the Bose-Einstein function $n(\omega)+1 = [1- \exp(-\hbar\omega/k_{B}T)]^{-1}$ ~\cite{Jaeckle}, which might suggest that mainly harmonic processes are involved. Nevertheless, recent experimental and theoretical work on perfectly ordered crystals ~\cite{Moratalla,Baggioli_PRL} has shown that the boson peak is a more general phenomenon which in certain circumstances may be generated by contributions of anharmonicity and optical modes.

A comparison between the VDOS of liquids (including INMs, from simulations) and that of glassy solids is shown in Fig. 1 to illustrate the main differences discussed above.

\begin{figure}
\includegraphics[height=5.7cm,width=9.0cm]{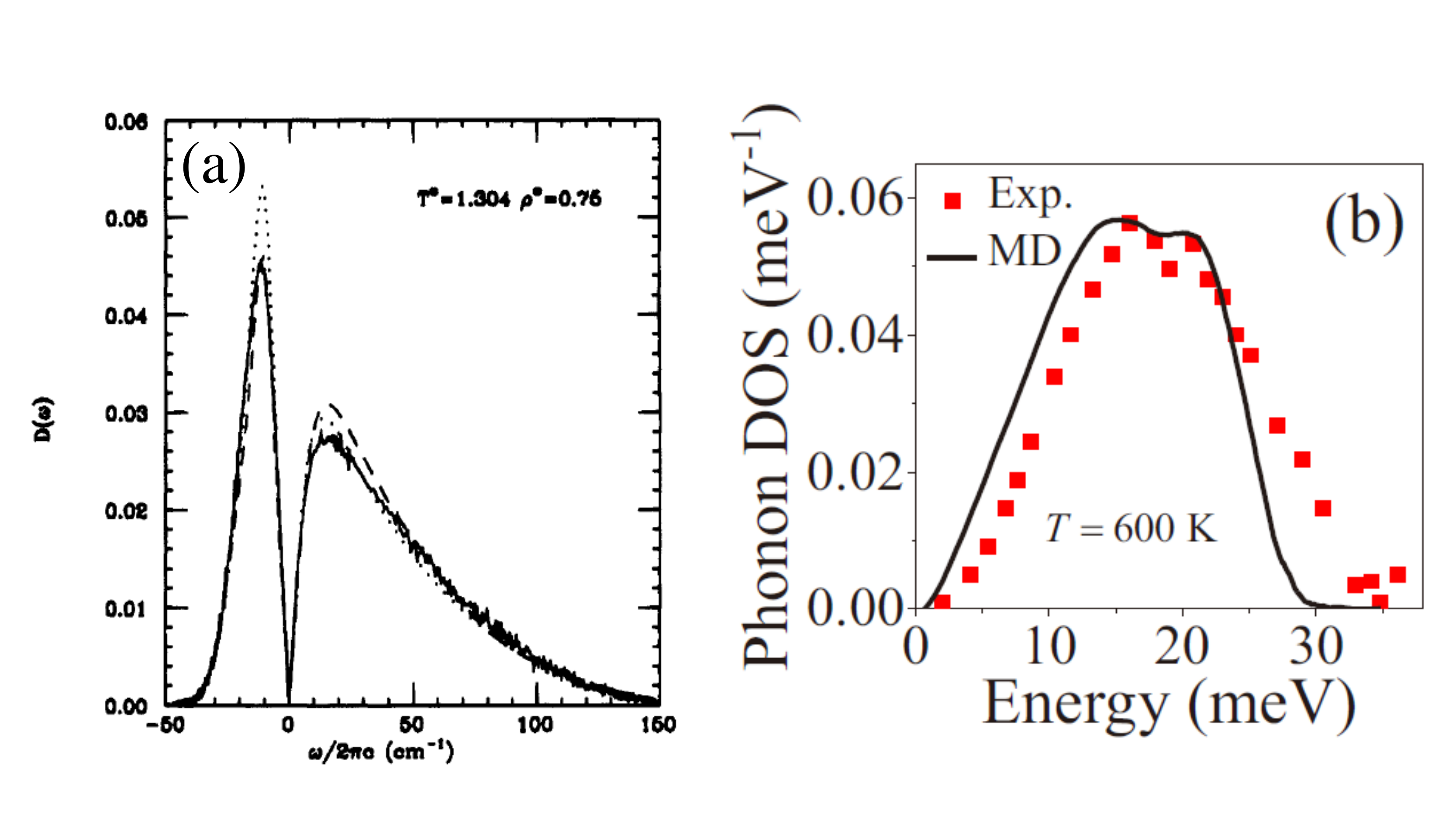}
\caption{(a) The vibrational density of states (VDOS) of a normal liquid, Argon, computed numerically and with theoretical approximations, the imaginary frequencies (instantaneous normal modes) corresponding to negative (unstable) eigenvalues of the Hessian are plotted on the negative axis. Reprinted (adapted) with permission from Ref. ~\cite{Stratt}. Copyright (2019) American Chemical Society. (b) The VDOS of a glassy solid below $T_g$, specifically the case of metal alloy $\textrm{Cu}_{50}\textrm{Zr}_{50}$, showing experimental data points (symbols) and MD simulations (solid line). Adapted from Ref.~\cite{Wang2019}. }
\end{figure}

\subsection{Linking relaxation and vibration}
Very few approaches have attempted to link relaxation and vibration in glassy systems. Historically, the first approach is rooted in Mode-Coupling theory (MCT), and provides a way to build the VDOS from the same dynamics of density fluctuations which controls the dynamical slow-down in MCT. This approach, as shown by Goetze and Mayr~\cite{Goetze_Mayr}, and also by Das~\cite{Das}, is able to recover the boson peak feature in the VDOS.

More recently, a different approach has been proposed for glasses below the $T_g$, which is based on lattice dynamics generalized to disordered solids. This approach, starting from the Newton equations for a disordered lattice of atoms/molecules in presence of viscous damping, arrives at a Generalized Langevin Equation in the particle coordinates which can be used also to describe the action of an external field and hence either dielectric or mechanical relaxation. Upon summing all contributions in eigenfrequency space, the VDOS naturally enters the description thus providing a direct link between relaxation and vibration. This approach will be described more in detail in Section VI.C.

Both the above approaches provide a link between the low-frequency features of the VDOS (in particular, the boson peak) and the $\alpha$-relaxation. 
However, there are at least three main open problems that have emerged over the years in the field of amorphous metal alloys where the relation between vibration and relaxation is crucial, yet it cannot be easily handled with current models, thus calling for new theoretical efforts.

The two main problems can be summarized as follows:\\
(i) the observation of compressed exponential relaxation with $\beta>1$ in glassy metal alloys below $T_g$;\\
(ii) the emergence of power-laws in both the vibrational spectrum as well as in the density correlations in metal alloys;\\
(iii) the microscopic nature of secondary ($\beta$) relaxation, in connection with local vibrational modes.

In the following sections, models of relaxation and vibration will be reviewed separately, and then models which attempt to establish a link between the two phenomena will be discussed more in detail. Finally, the open problems (i), (ii) and (iii) will be given detailed consideration along with ideas for future modelling strategies.

\section{Relaxation models}
A typical feature of glassy systems and not just of $\alpha$-relaxation, since its first observation by Kohlrausch in 1847~\cite{Cardona}, has been observed in the time-dependent relaxation
of various (viscoeelastic, dielectric, electrical) macroscopic observables in nearly all structurally disordered solids and in crystalline solids with defects. Stretched-exponential relaxation has recently been observed also in the resistivity of thermally-annealed indium oxide films~\cite{Ovadyahu}.
Over the last century, stretched-exponentials have been used in countless experimental contexts to provide empirical fittings to experimental data. Although it is common knowledge that
stretched-exponential relaxation relates somehow to spatially heterogeneous many-body interactions or to heterogeneous distribution of activation energy barriers
~\cite{Montroll,Langer}, only very few models or theories are able to predict stretched-exponential relaxation from first-principle dynamics~\cite{Phillips}. 

There are at least three main theoretical frameworks which are able to explain and recover stretched-exponential relaxation starting from very different assumptions, which are reviewed below. 

\subsection{The trap model}
This is a model of electronic relaxation via non-radiative exciton-hole recombination where holes are randomly distributed traps that "eat up" the diffusing excitons, originally proposed by Ilya M. Lifshitz and represented in later works in particular by Procaccia and Grassberger, and by J.C. Phillips~\cite{Lifshitz,Procaccia,Phillips}. As shown in Ref.~\cite{Procaccia}, according to this trap model, under the assumptions of diffusive dynamics with an annihilation term and Poisson statistics, one finds that the density of not-yet trapped (or survival probability) of excitons decays at long times as $\sim \exp(-t^{d/d+2})$, where $d$ is the space dimensionality, which gives a Kohlrausch exponent $\beta=0.6$ in $d=3$. In spite of the elegance of this model, it is not straightforward to apply it to elucidate Kohlrausch relaxation in supercooled liquids and structural glasses, due to the difficulty of identifying proper excitations and traps in those systems. Nevertheless, the trap model  predicts ''magic values" of the stretching exponent $\beta=3/5$, from putting $d=3$ in $\sim \exp(-t^{d/d+2})$, and $\beta=3/7$ for systems with long-range relaxation where long-range relaxation channels are characterized by an effective fractal dimensionality $d=3/2$ leading to $\beta=3/7$. These values found experimental confirmation in some systems~\cite{Mauro} and may be interpreted in terms of effective dimensionalities in molecular configuration space.
In supercooled liquids, however, it rather appears that there is a continuous spectrum of $\beta$ values which continuously evolves (increases) with temperature, instead of discrete "magic" values, as suggested by experimental dielectric measurements~\cite{Casalini2003}.

\subsection{Trachenko's cage rearrangement model for stretched-exponential relaxation}
Trachenko~\cite{Trachenko2007} proposed a model based on the concept of Local Relaxation Events (LREs), which may be described as rearrangements of the nearest-neighbour cage whereby an atom escapes from the cage. The escape process is associated with a mechanical distortion of the cage, hence with an additional stress. This relaxes the local stress, resulting in more local stresses experienced by other cages in the system. Within a scheme similar to the elastic or shoving model~\cite{Dyre_1998}, this additional stress times the local activation volume (roughly, the volume of the cage) causes an increases of the energy barrier for the relaxation events that are yet to occur. As time goes by, more relaxation events take place (ageing), such that the energy barrier keeps increasing, through a feedforward mechanism. Upon setting a kinetic equation for the time derivative of the number of LREs as proportional to the number of LREs that still need to take place (assuming there is a fixed total number of LREs at time equal infinity) and also proportional to an Arrhenius factor of the energy barrier (the relaxation event probability), the resulting solutions were studied. If the energy barrier in the Arrhenius factor is constant with time, simple exponential relaxation is recovered. If, instead, the energy barrier increases with the number of LREs, the feedforward mechanism leads to stretched-exponential relaxation. Interestingly, the value of stretching exponent $\beta$ in this model decreases upon decreasing $T$ in a predictable way, consistent with most experimental measurements.

\subsection{Mode-Coupling Theory}
The other widespread theoretical framework is the Mode-Coupling Theory (MCT) of supercooled liquids. This is a radically different approach based on hydrodynamics, where atomic dynamics on the Liouvillian level is systematically coarse-grained by means of projection-operator techniques. By means of this coarse-graining process in phase space, the slow variables (density fluctuations) can be isolated and shown to satisfy a self-consistent Generalized Langevin Equation (GLE) for the dynamics, where, typically, the dependent variable is given by  the intermediate scattering function $F(q,t)$ where $q$ is the wavevector (note that in the following we will use $q$ and $k$ for the wavevector interchangeably). The damping or frictional term in the GLE is represented by a non-Markovian memory kernel, which itself is a function of $F(q,t)$, thus making the GLE self-consistent and to be solved numerically~\cite{Goetze_1992, Goetze, Mazenko}. 

Much research has been devoted to determining the memory kernel for the damping term in the GLE-like dynamical equation of MCT, which normally requires a number of assumptions. The theory can then be shown to produce a non-ergodicity transition at a dynamical crossover temperature, below which the $F(q,t)$ does not decay to zero at long time, but reaches a plateau (localization or caging). This caging transition occurs at temperatures that are somewhat higher than the actual glass transition temperature of the system. In the $\alpha$-relaxation regime, the decay of $F(q,t)$ at wavevectors comparable to the size of a nearest-neighbour cage can be fitted with a stretched-exponential function~\cite{Goetze_1999,Goetze_1992}. In general, as emphasized by G{\"o}tze and co-workers, no special meaning can be attributed to the value of $\beta$ since this is essentially the result of a fitting procedure. However, there is an asymptotic limit of MCT where the stretched-exponential relaxation becomes exact, namely at $q\to\infty$, as was pointed out in Ref.~\cite{Fuchs1994}.
On time-scales shorter than the $\alpha$-relaxation process, MCT successfully predicts power-law decay in time including the von Schweidler relaxation  which also shows up as a power-law in frequency behaviour in the high-frequency spectrum of dielectric relaxation~\cite{Goetze}.

\subsection{Random energy landscape (dynamical heterogeneity)}
Yet a different way of recovering stretched-exponential relaxation has been studied in the field of spin glasses. Such systems can have a number of quasi-degenerate energy states or minima in the energy landscape. These energies can be treated as random independent variables, as is done in the analytically-solvable random energy model (REM) developed by Derrida~\cite{Derrida}. The relaxation to equilibrium of these models has been studied within the frame of a master kinetic equation by De Dominicis, Orland and Lainee~\cite{DeDominicis}. These authors found that the self-correlation for the probability of finding the system in a state with a certain energy behaves like a stretched-exponential at long time. 
Also this result is derived under equilibrium-like assumptions such as imposing detailed-balance in order to solve the master kinetic equation, and thus hardly applicable to frozen-in glassy systems that are out of equilibrium. 

More generically, and also more simplistically, one can describe the energy landscape of a glassy system as a rugged function with many minima separated from each other by barriers (saddles) of variable height $E$.
This naturally introduces a distribution of barrier heights. If the main transport mechanism is diffusive, relaxation from a well into another across the barrier is a simple exponential process $\sim \exp(-t/\tau)$. Since there is a distribution of relaxation times $\rho(\tau)$ (induced by the distribution of barrier heights), the global correlation function will be a weighted average of the form~\cite{Anderson1984}
\begin{equation}
\phi(t) \sim \int_{0}^{\infty} \rho(\tau) \exp(-t/\tau) d\tau 
\end{equation}
if $\rho(\tau)$ is Gaussian, it is very easy to see that $\phi(t)$ starts off as a simple exponential decay and then crosses over into a stretched-exponential with $\beta \approx 2/3$. This is a typical value of the stretching exponent that is observed in the $\alpha$-relaxation of structural glasses. 
A similar result can be obtained assuming that the quenched disorder is in the rate of jumping of an atom outside of the nearest-neighbour cage, and has been numerically demonstrated in Ref.~\cite{Rabani}. 
An assumption underlying Eq. (1) is that the relaxation spectrum $\rho(\tau)$ exists \textit{a priori} and does not evolve with time on the observation time scale. 

In general, a stretched-exponential relaxation will result from the integral in Eq. (1) if the distribution $\rho(\tau)$ satisfies certain (merely mathematical) properties. 
The properties of the distribution (including its moments) can be linked to the value of $\beta$ as discussed in Ref. ~\cite{Johnston}. 

In general, while stretched-exponential relaxation is ubiquitous in the solid-state, it has not been possible to trace it back to a single well-defined mechanism in the many-body dynamics, or to a well-defined microscopic descriptor of the dynamics so that this remains a very active field of research.

\subsection{$\beta$ relaxation}
Secondary or $\beta$ relaxation typically manifests itself as a secondary (smaller) peak in the dielectric loss $\epsilon''$ as a function of electric field frequency~\cite{PLunk2000}, or in the viscoelastic loss modulus as a function of the mechanical oscillation frequency. It can also be inferred from the viscoelastic loss modulus measured as a function of temperature, as was done for metallic glasses in Ref.~\cite{Roesner} where it shows up as a low-T wing attached to the $\alpha$-peak. Initially discovered by Johari and Goldstein in molecular glass formers, it was demonstrated to be a feature not only of molecules with excitable internal degrees of freedom, but also present in liquids/glasses formed by rigid molecules~\cite{Johari-Goldstein}. From a theoretical point of view, thanks to work done within the frame of MCT, $\beta$ relaxation can be understood as a faster relaxation decay which precedes $\alpha$-relaxation~\cite{Goetze_1992} and is described, within MCT and in agreement with simulations and experiments, as a power-law relaxation in time. 
Furthermore, in metallic alloys, an analysis based on the so-called coupling model, which establishes a link between the characteristic time scale of the Johari-Goldstein relaxation and that of the $\alpha$-relaxation, has shown that the experimentally~\cite{Roesner} estimated time scale of $\beta$ relaxation does indeed coincide with the time scale of a genuine Johari-Goldstein process~\cite{Ngai2006}.

Other approaches, in particular simulations, have highlighted the prominence of mesoscopic cluster motions, in particular string-like motions, in the $\beta$ relaxation process~\cite{Douglas}, a phenomenon that has been observed also in metal alloys~\cite{Samwer,Samwer_review,Douglas_metal}, and from a theoretical point of view can be recovered from the Random First Order Theory (RFOT) of the glass transition upon considering the configurational entropy of the clusters~\cite{Wolynes_beta}. 

In the current state of the art, we therefore have, on one hand, theoretical models, such as MCT, which can provide a good description of the relaxation dynamics for $\beta$  relaxation but fail to capture the microscopic nature of thermally-activated jumps and locally cooperative motions involved in the process. Instead, simulations, on the other hand, have pinned down the microscopic nature of the motions involved in the relaxation process but do not provide analytical insights into the time dependence of correlations functions. 

A further promising line of research is based on the energy landscape concept~\cite{Johari_energy}, and has recently brought new insights. In particular, it has been recognized that the $\beta$ relaxation process kinetics is related to the rate of jumps from one local minimum to an adjacent one, within the same meta-basin of the energy landscape. In higher dimensions, it has been recently recognized that the structure of the local minima becomes self-similar or fractal beyond a certain critical temperature (or a critical value of some other control parameter), a transition called the Gardner transition in analogy with similar phenomena in spin glasses~\cite{Kirkpatrick,Zamponi_fractal}. This approach may, in future work, bring a more complete picture of the secondary relaxation process by elucidating the connection between relaxation dynamics/kinetics and the localized force-network and vibrational modes associated with secondary relaxation. 

Finally, orientationally-disordered crystals (or plastic crystals) have provided a new playground with new opportunities to precisely isolate molecular motions associated with $\alpha$ and $\beta$ relaxation. In particular, it has been shown using Nuclear Magnetic Quadrupole resonance techniques~\cite{Zuriaga2009} that $\alpha$ and $\beta$ relaxation may arise from molecular re-orientational motions with different dynamics that can be ascribed, separately, to molecules in the crystalline lattice that belong to non-equivalent molecular environments from the point of view of symmetry. These findings highlight the need to reconsider $\alpha$ and $\beta$ relaxation phenomena in the perspective of the lack of a universal microscopic mechanism.

\section{Models of vibrational spectra}
Early research into the vibrational spectrum and the VDOS of glassy solids has been motivated by the discovery, by Zeller and Pohl in the early 1970s, of certain anomalies in the specific heat and in the thermal conductivity of glasses at low temperatures. In particular, it was realized that the specific heat of glasses at low T deviates from the $T^{3}$ predicted by the Debye model, and exhibits instead a linear in $T$ behaviour. Similarly, the thermal conductivity behaves like $T^{2}$, instead of $T^{3}$ predicted again from the Debye model together with Peierls-Boltzmann transport assumptions. 
Clearly, since the VDOS enters as a factor inside the integrals over eigenfrequency space which define the specific heat and the thermal conductivity, any deviations from the $\sim \omega^{2}$ law are expected to lead to deviations from the $T^{3}$ scaling, which represented a motivation to study and understand the vibrational spectra of glasses.
For example, for the specific heat, the relationship to the VDOS is given by the following expression:
\begin{equation}
C(T)\,=\,k_B\,\int_0^\infty\, \left(\frac{\hbar \omega}{2\,k_B\,T}\right)^2\,\sinh \left(\frac{\hbar \omega}{2\,k_B\,T}\right)^{-2}\,D(\omega)\,d\omega
\end{equation}
where $D(\omega)$ is the VDOS.

As a model system, the silica glass has been used in early experimental studies of the VDOS of glasses~\cite{Buchenau1986} using INS which provided early evidence of the excess of vibrational states above the $\sim \omega^{2}$ prediction, known as the boson peak. 

Since then, the boson peak feature has been observed in practically all glassy materials: polymer glasses~\cite{Sokolov}, amorphous metal alloys~\cite{Evenson,WangPRL}, colloidal glasses~\cite{Bonn}, jammed packings~\cite{O'Hern}, proteins~\cite{Sokolov_protein}, orientationally disordered organic crystals~\cite{SzewczykTJPC2015}, and the already mentioned oxides~\cite{Monaco}, and several theoretical approaches have been developed to describe this phenomenon. Below we briefly review some of the theoretical models proposed for the description of the VDOS of glasses.

\subsection{Fluctuating elasticity (FE) model}
One of the most popular approaches is based on evaluating the Green's function within a self-consistent Born approximation assuming that the shear modulus is spatially heterogeneous, i.e. the fluctuating elasticity (FE) model by Schirmacher, Ruocco and co-workers~\cite{Schirmacher}. 
The starting point is typically the assumption that the shear elastic modulus fluctuates throughout the material around an average value, with Gaussian statistics (although the latter specification is not necessary, other non-Gaussian distributions lead to similar results). An elastic Lagrangian is then formulated and the Gaussian disorder assumption leads straightforward to the evaluation of the free energy with quenched disorder and hence to the evaluation of the Green's function. In the simplest version of the theory, the Green's function resembles the Green's function of a damped harmonic oscillator~\cite{Schirmacher_Maurer}, 
\begin{equation}
G(z)=\sum_{\mid k \mid<k_D}G(k,z)=\sum_{\mid k \mid<k_D}\frac{1}{-z^{2}+k^{2}[c^{2}-\Sigma(z)]}
\end{equation}
where $z=\omega+i\epsilon$, with $\epsilon\rightarrow 0$, $c$ is the speed of sound, while $\Sigma$ is the self-energy which is self-consistently coupled to $G(z)$ (resulting from resummation of the Dyson series), and contains the effect of multiple scattering of the phonons by the disorder in the elastic modulus.
From the Green's function, the VDOS can be computed using the Plemelj identity which leads to:
\begin{equation}
D(\omega)=\frac{2\omega}{\pi} \textrm{Im} G(z).
\end{equation}

The self-energy $\Sigma(z) $contains information about scattering processes induced by the disorder. Hence the phonon damping associated with the self-energy is a genuine manifestation of "harmonic", or dissipationless, disorder. 

The same Green's function evaluated from the FE model can be used to compute the viscoelastic moduli $G'$ and $G''$ as a function of shear oscillation frequency~\cite{Schirmacher_viscoelastic}. The resulting framework is able to recover the $\alpha$-relaxation peak in the loss viscoelastic modulus $G''$, see the Section VI.B for a more detailed discussion.

This model has also the historical merit of unveiling the deep connection between vibrational modes that constitute the boson peak and the prominence of transverse modes associated with shear elasticity.

\subsection{Soft potential model}
Historically the soft potential model has been the first theoretical model that aimed to explain the VDOS of glasses. The main idea is that randomly-distributed defects live in a highly anharmonic environment, and experience soft mode-type vibrations which  add up onto the Debye phonons thus resulting in the boson peak excess of vibrational states~\cite{Gurevich1993,Parshin_1993,Klinger}. 
The key assumption of anharmonicity has been shown in recent work to be inessential to capture the essence of the boson peak. However, the soft potential model provided the first prediction of the existence of quasi-localized modes at frequencies lower than the boson peak, the distribution of which locally goes like $\omega^{4}$. These modes have been discovered numerically in recent molecular simulations where careful tuning of the system sizes allows one to disentangle these $\omega^{4}$ modes from the acoustic phonons~\cite{Lerner1,Lerner2}. 
An independent $1D$ derivation of these modes based on a simple localization model was provided by Gurarie and Chalker~\cite{Gurarie}. 

\subsection{Shifted/smeared van Hove singularity}
An important model to explain the occurrence of the boson peak in glasses and semi-ordered system has been the one based on adding disordered perturbatively to the lattice dynamics in order to study the shifting and broadening of van Hove (vH) singularities. Within this model the boson peak can be identified with a shifted/smeared out vH singularity. We recall that a vH singularity occurs whenever the phonon dispersion relation $\omega(q)$ becomes flat with $q$, i.e. at a (pseudo) Brillouin zone (BZ) boundary. Since the VDOS can be expressed as an integral where the $q$-gradient of the dispersion relation appears in the denominator, the flattening of the dispersion relation at the BZ boundary clearly leads locally to a singularity. 
This features as a sharp peak in the VDOS of crystals. The presence of disorder has been shown to cause a lowering and broadening of the van Hove singularity. 
Recent works~\cite{Schirmacher_vanHove,Albe} have questioned the prominence of this effect in determining the boson peak. 
Also, the analysis of the VDOS of harmonic FCC lattices where disorder can be tuned have shown that the boson peak and the smeared-out lowest van Hove singularity occur as two well separated peaks in the VDOS, with the boson peak showing up at significantly lower frequency than the smeared-out vH singularity~\cite{Milkus}. 

\subsection{Nonaffine/inversion-symmetry model}
The boson peak has been put in relation with the breaking down of continuum elasticity in glasses on mesoscopic lengthscales~\cite{Tanguy1,Tanguy2}. This implies that the atomic-level deformation is different from the macroscopic deformation controlled by the macroscopic strain tensor, which leads to a softening of the shear modulus of the amorphous glass with respect to its crystalline fully-ordered counterpart (for same density, bonding etc). Growing evidence has pointed towards a prominent role of transverse (shear) modes which  cross over from ballistic propagation to quasi-localized diffusive transport (diffusons) at a Ioffe-Regel frequency close to the boson peak frequency~\cite{Parshin}. Furthermore, the height of the boson peak has been shown, in 2D  simulations, to scale as $\sim 1/G$, where $G$ is the shear modulus. 

\begin{figure}
\includegraphics[height=5.8cm,width=9.0cm]{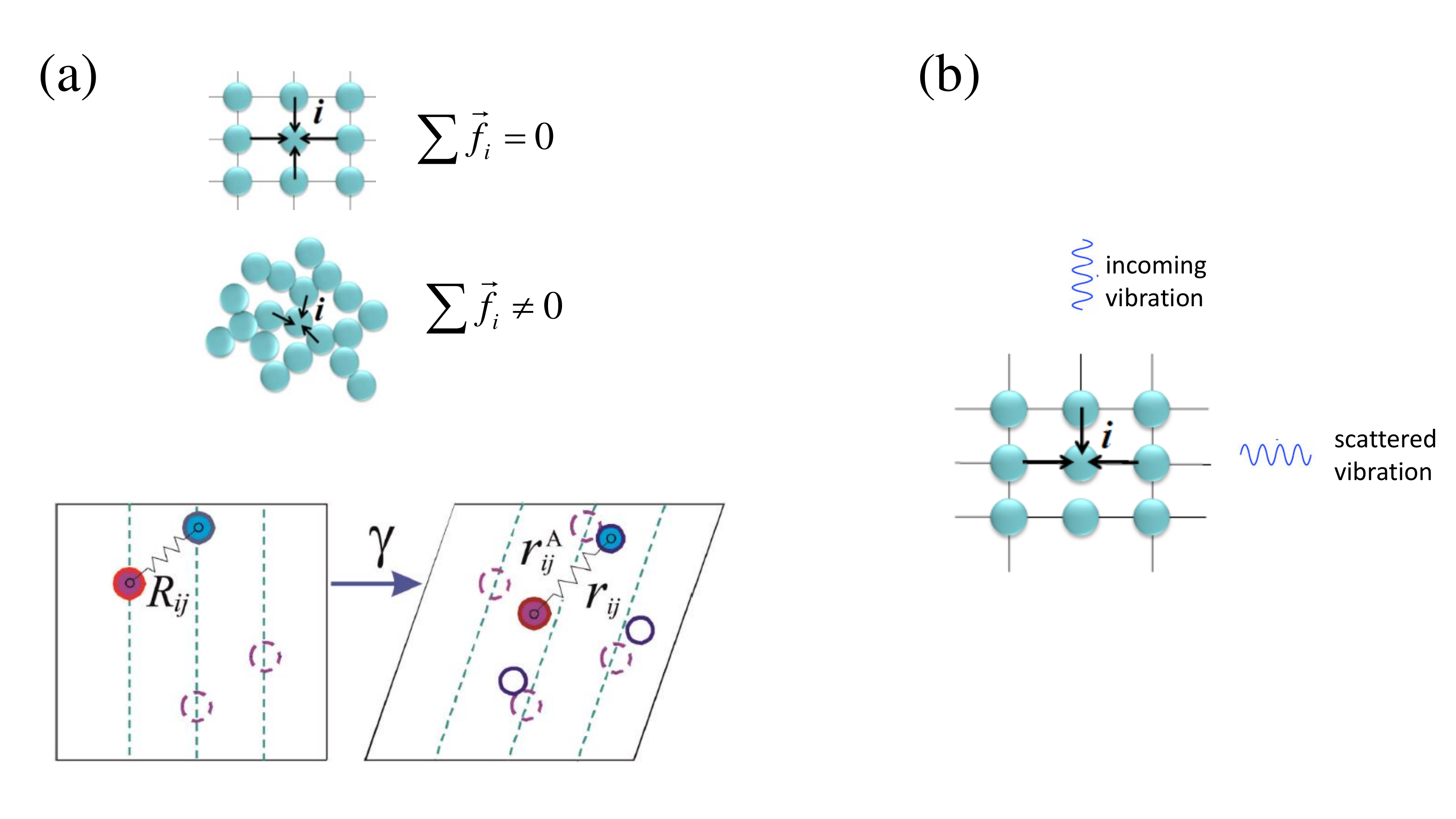}
\caption{(a) The top cartoon shows a comparison of the force-balance in the affine position (prescribed by the macroscopic strain tensor) in a centrosymmetric crystal (top) and in a glass (bottom). In the affine position in the centrosymmetric crystal all forces from nearest neighbours cancel by symmetry and the atom $i$ is automatically in mechanical equilibrium, no further displacement is needed. In the glass, the forces do not cancel due to the lack of inversion symmetry, hence there is a net non-zero force in the affine position that has to be relaxed through an additional \textit{nonaffine} displacement. The bottom cartoon depicts the nonaffine displacements for two bonded atoms: if the displacements were purely affine (like in a centrosymmetric crystal) the atoms would still lie on the dashed lines also in the deformed (sheared) configuration. Instead, in glasses, they are to be found away from the dashed line, and the nonaffine displacements are defined as the distance between the actual end position (away from the dashed line) and the affine position (on the dashed line). $R_{ij}$ indicates the interatomic distance at rest, whereas $r_{ij}^{A}$ labels the interatomic distance if the atoms were both in the affine positions. (b) Schematic representation of what happens in a defective (disordered) crystal lattice e.g. when bonds are randomly removed, or, which is the same, in the presence of vacancies. An incoming phonon cannot continue along the original direction due to the missing bond and it is therefore deviated (scattered) into a different direction. This illustrates the role of non-centrosymmetry (in either atomic positions or atomic bonding) in promoting scattering phenomena on mesoscopic length scales which in turn cause a crossover from ballistic phonon propagation into diffusive-like transport due to scattering where vibrational excitations become quasi-localized (Ioffe-Regel crossover). The same phenomenology is likely to be at play in glasses and defective crystals~\cite{Milkus}. }
\end{figure}

In parallel, much work has been devoted to reformulating lattice dynamics in order to account for disorder. These efforts, which started with early molecular dynamics work aiming to account for atomic relaxations in the deformation of solids~\cite{Lutsko}, have led to an extended lattice dynamics formalism called nonaffine lattice dynamics (NALD)~\cite{Lemaitre,Scossa-Romano}. The basic idea is that atoms in an amorphous solid do not occupy centers of inversion symmetry. As a consequence, as atoms move towards their prescribed positions (imposed by the applied strain tensor) which are called affine positions, they are subject to forces from their nearest-neighbours (also moving towards their respective positions). In a centrosymmetric crystal lattice all these forces acting on atom $i$ would cancel each other out by inversion symmetry.
Instead, in an amorphous solid (or in a non-centrosymmetric lattice such as $\alpha$-quartz) these forces cannot identically sum up to zero, and a residual force acts on atom $i$ in the affine position. The only way to relax this force is that atom $i$ undergoes an additional displacement away from the affine position. This additional displacement is called \textit{nonaffine} displacement, and plays the key role in the physics of deformation of glassy materials, in particular the nonaffine displacement times the force that acts on the atom provides a negative contribution to the internal energy of deformation of the solid, which results in the mechanical weaking.
The nonaffine deformation mechanism is depicted in the cartoon in Fig. 2(a). 

The local breaking of inversion symmetry is not only responsible for the mechanical softening of the material, but also affects phonon transport, as schematically depicted in Fig.2(b). In recent work Ref.~\cite{Milkus}, it has been shown on the paradigmatic example of harmonic disordered lattices that the boson peak height strongly correlates with an order parameter that quantifies the local degree of inversion-symmetry breaking. It does not correlate at all, instead, with the usual bond-orientational order parameter.
In other words, the stronger the statistical lack of inversion-symmetry in the atomic environment, the stronger the boson peak. 
This picture is thus consistent with identifying the local degree of non-centrosymmetry as a key atomistic structural feature that promotes the scattering of phonons on mesoscopic length-scales, as depicted schematically for the case of a defective crystal lattice in Fig. 2(b). This also provides a unifying framework for both glasses and disordered crystals. 

\subsection{The fracton model}
Alexander and Orbach~\cite{Alexander-Orbach} famously proposed a simple model for the density of states on a fractal, called the fracton model or the Alexander-Orbach model.
Their starting point was a relationship between the DOS and the single site Green's function, similar to Eq. 4 above, written for the spectral eigenvalue $\lambda$, 
\begin{equation}
\rho(\lambda)=-\frac{1}{\pi}\textrm{Im}\langle \tilde{P}_{0}(-\lambda +i0^{+})\rangle
\end{equation}
where $\tilde{P}_{0}$ is the Laplace transform of the time-correlation function $P_{0} (t)$ which tells the position of a particle (initially at the origin) after a time $t$, and is clearly related to the Green's function $G(z)$ introduced above. Assuming anomalous or fractional diffusion as appropriate for transport on a fractal object, the mean square displacement reads as $\langle r^{2}(t) \rangle \sim t^{2/(2+\bar{\delta})}$, where $\bar{\delta}$ is the anomalous diffusion index which depends on the geometry. Assuming furthermore that $\langle P_0 (t) \rangle \sim [V(t)]^{-1}$, where $V(t) \sim \langle r^{2}(t) \rangle^{\bar{d}/2}$ denotes the available volume for the diffusion process with $\bar{d}$ the fractal dimension of the object, they obtained $\langle P_0 (t) \rangle \sim t^{-\bar{d}/(2+\bar{\delta})}$. Upon replacing the latter result in Eq. (3) they arrived at the following result for the VDOS:
\begin{equation}
N(\lambda) \sim \lambda^{[\bar{d}/(2+\bar{\delta})]-1}
\end{equation}
and upon changing to eigenfrequency:
\begin{equation}
D(\omega) \sim \omega^{[2\bar{d}/(2+\bar{\delta})]-1}
\end{equation}

This model thus produces a VDOS which also departs from the Debye prediction $\sim \omega^{2}$, on account of the fractal structure of the material. Furthermore, it has been shown to produce a Ioffe-Regel crossover from propagating phonons to (quasi)localized excitations (having a mean-free path lower than their wavelength) called \textit{fractons}~\cite{Aharony}, where the localization crossover is related to the existence of a correlation length in the self-similar structure. This model will be used in the following to interpret emerging power-laws in the vibrational and relaxation properties of metal alloys.

\section{Emerging picture for the VDOS of disordered solids}
Growing consensus is pointing towards a Ioffe-Regel crossover from propagating (Debye) phonons to quasi-localized ''diffusons", controlled by structural disorder, as the origin of the boson peak~\cite{Ruffle,Tanaka,Schirmacher_SciRep,Schirmacher_PSSB,Zamponi,Parshin,Mizuno_Barrat,Mizuno,Milkus,Manning,Baggioli_random}. In particular, the frequency of the crossover, which controls the frequency of the boson peak, can be identified with the frequency at which phonons start to feel the random-matrix structure of the eigenvalues of the Hessian (dynamical) matrix. 

The eigenvalue distribution of a sparse random matrix with stochastic entries tends to follow a simple analytical law known as the Marchenko-Pastur (MP) distribution, which is a generalization of the Wigner semi-circle law. In particular, random matrices drawn from the Wishart ensemble of matrices $M$ have a distribution of eigenvalues given by the MP distribution. The Wishart ensemble is created by starting from a $m \times n$ Gaussian random matrix $ A$ using $
	 M \,=\, \frac{1}{n}  A \, A^T $. This matrix has the eigenvalue distribution ($ M\, v 
	\,=\, \lambda v $) :
	\begin{equation}
	p(\lambda)\,=\,\frac{\sqrt{((1+\sqrt{\rho})^2-\lambda)(\lambda - (1-\sqrt{\rho})^2)}}{2 \pi \rho \lambda}\quad
	\end{equation}
	where we introduced the parameter $\rho = m/n$. Since we are interested in the vibrational density of states $D(\omega)$ of the eigenfrequencies $\omega = \sqrt{\lambda}$ we transform $p(\lambda)$ to the frequency space: $p(\lambda) d\lambda = D(\omega) d\omega$
	\begin{equation}
	D_{random}(\omega)\,=\,\frac{\sqrt{((1+\sqrt{\rho})^2-\omega^2)(\omega^2 - (1-\sqrt{\rho})^2)}}{\pi \rho\, \omega}
	\end{equation}
This contribution to the VDOS represents the effect of randomness in the Hessian, and still does not contain features related to the existence of Debye phonons and of short-range breaking of translation invariance  reponsible (also in glasses) for van Hove-like peaks. These effects can be implemented in the following way.	
	
	In order to accommodate phonons, one can shift this distribution in the frequency space by $\delta$ and introduce the width of the spectrum $b$ by the transformation:	$\omega\,\rightarrow\,\frac{2}{b}\,\left(\omega\,-\,\delta\right)$,
which gives:
\begin{widetext}
	\begin{equation}
	D(\omega)=\frac{\sqrt{((1+\sqrt{\rho})^2-\frac{4}{b^2}(\omega-\delta)^2)(\frac{4}{b^2}(\omega-\delta)^2 - (1-\sqrt{\rho})^2)}}{\pi \rho \frac{2}{b}|\omega-\delta|}.
	\end{equation}
\end{widetext}
This shifted spectrum belongs to a matrix $M'$ that can be derived by the original Wishart matrix $ M$ in the following way~\cite{ParshinJETP,Parshin}: $(M')^{1/2} \,=\,\frac{b}{2}  M^{1/2} + \delta \textbf{1}$,
where $\delta$ and $b$ both depend on the minimal and maximal eigenfrequencies of the system, $\omega_{-}=\frac{b}{2}(1-\sqrt{\rho})+\delta$ and $\omega_{+}=\frac{b}{2}(1+\sqrt{\rho})+\delta$, respectively, which define the support of the random matrix spectrum. 

In particular, the value of $\delta$ controls the shift of the lower extremum of the support of the random matrix distribution, and thus its value controls the frequency $\omega^* \approx \omega_{-}$ which is associated with the boson peak. 

We still need to make two additional modifications that cannot be induced by a corresponding change in the matrix $ M'$: first we need to correct the lowest edge of the spectrum by a factor that behaves like $\sim \omega^{-1/2}$ for $\omega \rightarrow 0$ and like $\sim 1$ for $\omega \gg 0$, and second we need to add peak functions to model the relics of the van Hove singularity peaks which become more prominent for systems with high values of $Z$ due to the topology of the random network becoming influenced by the limiting FCC lattice to which any lattice will converge for $Z=12$. This second correction is achieved by modelling the two relics of the van Hove peaks with two Gaussian functions. The final result for the fitting formulae of the VDOS reads~\cite{Baggioli_random}: 	
	\begin{align}
D(\omega)=&1.2\frac{\sqrt{[(1+\sqrt{\rho})^2-\frac{4}{b^2}(\omega-\delta)^2][\frac{4}{b^2}(\omega-\delta)^2 - (1-\sqrt{\rho})^2]}}{\pi \rho \frac{2}{b}|\omega-\delta|}\notag\\
	&\times \left(\left(\frac{0.65}{\omega}\right)^2+0.25 \right)^{1/4} + G_1(Z,\omega) + G_2(Z,\omega)
\end{align}
where $G_1= (0.011(Z - 6)^2 + 0.175)\sqrt{\frac{2}{\pi}}\exp(-2 (\omega - 1.6)^2)$ and $G_2=(0.011(Z - 6) + 0.045)\sqrt{\frac{8}{\pi}}\exp(-8 (\omega - 2.3 - 0.07(Z-6))^2)$ are the two Gaussian functions used to model the van Hove peaks. The parameters of these Gaussian functions are chosen empirically to match the numerical data, although the centers of the Gaussians must be related to the pseudo-Brillouin zone boundaries of the systems. The latter must coincide with those of the FCC crystal which one recovers in the limit $Z=12$.

Equation (11) with $\rho=1.6$ can now be compared with numerical data obtained for random networks of harmonic springs which have a narrow distribution of spring length and all have the same spring constant. These networks were obtained from a precursor Lennard-Jones glass upon replacing nearest-neigibhour interactions with harmonic springs (and removing further than nearest neighbour interactions)~\cite{Milkus}. The disorder in the numerically generated networks can be tuned upon randomly removing springs, all the way from the FCC limit $Z=12$ (ordered crystal) down to the marginally-rigid system with $Z=6$ (the "epitome" of disorder). 
The comparison is shown in Fig. 3. In particular, the expressions of all the parameters $\delta$, $b$, $\rho$, and those inside $G_1$, $G_2$, either remain fixed upon changing $Z$ or evolve with $Z$. Hence, Eq. (11) captures the variation of the VDOS  spectrum upon varying the coordination number $Z$ of the network.

The model parametrization given by Eq. (11) is excellent and provides an accurate description of the data for all $Z$ values considered in the broad range from $Z=9$ down to the unjamming transition at $Z=6$. In particular, the expressions of all the parameters $\delta$, $b$, $\rho$, and those inside $G_1$, $G_2$, either remain fixed upon changing $Z$ or evolve with $Z$. Hence, Eq. (11) captures the variation of the VDOS  spectrum upon varying the coordination number $Z$ of the network.

\begin{figure}
\centering
\includegraphics[width=.85\linewidth]{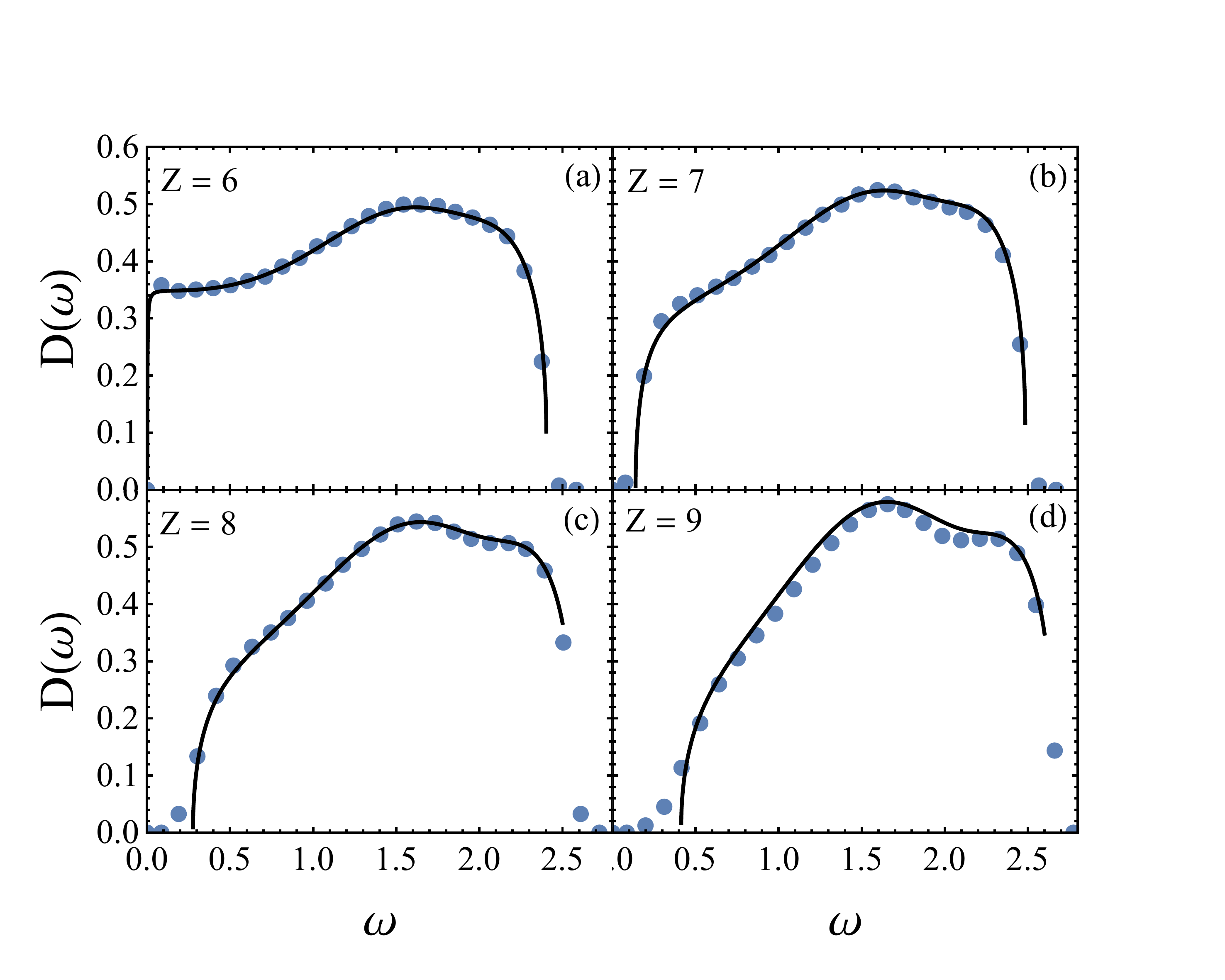}
\caption{Numerical VDOS (symbols) of harmonic random networs and random-matrix model analytical fitting provided by Eq. (11) (solid lines)~\cite{Baggioli_random}. Debye phonons are visible (as quadratically growing from $\omega=0$) in the numerical data, but are not implemented in the analytical model.}
\label{fig2}
\end{figure}

A word of caution here is in order. The random-matrix model presented above (and originally developed in ~\cite{Baggioli_random}), is based on extrapolating the MP distribution which is exact for an infinite-dimensional space (see e.g. the discussion in ~\cite{Zamponi} and in ~\cite{Cicuta}) to a 3d lattice. This means that, in the reality of a 3d atomic lattice in a solid, the Hessian matrix is a block random matrix, with  $3 \times 3$ blocks due to the fact that each interatomic connection is described by a unit vector $\textbf{n}_{ij}$, in 3d. The Hessian matrix for a harmonic random network reads as:
\begin{equation}
H_{ij}^{\alpha\beta}= \delta_{ij}\sum_{s}\kappa c_{is} n_{is}^\alpha
n_{is}^\beta -(1-\delta_{ij}) \kappa c_{ij} n_{ij}^\alpha
n_{ij}^\beta\label{eqmod9}
\end{equation}
where $\alpha, \beta$ are Cartesian indices, $c_{ij}$ is a random (scalar) coefficient matrix with $c_{ij}=1$ if atoms $i$ and $j$ are nearest neighbors and $c_{ij}=0$ otherwise. $c_{ij}$ is a matrix where each row and each column have on average $z$ elements equal to 1 distributed randomly with the constraint that the matrix be symmetric.
Hence, each $ij$ entry of the Hessian is a $3 \times 3$ block given by the factor $n_{ij}^\alpha
n_{ij}^\beta$. This is not taken into account by the MP distribution, which has been shown in Ref.~\cite{Cicuta} to converge to the actual distribution of the Hessian only in the limit of $d \rightarrow \infty$. This is a result consistent with independent results obtained within the context of the replica theory of glasses in Ref.~\cite{Zamponi}. The eigenvalue spectrum of the actual 3d Hessian can only be obtained numerically, but it has been shown that the differences with the MP-derived spectrum, although noticeable, are not essential, especially in the regime of the boson peak. 

Hence, the above comparison provides a strong indication of how the VDOS in the vicinity of the boson peak, i.e. in the region where one sees the crossover from Debye phonons to random-matrix behavior, is effectively dominated by the random-matrix statistics of the Hessian's eigenvalues.

Future work should be addressed to provide further insights into the deeper relation between random-matrix behaviour of the VDOS and the quasi-localized diffusons nature of vibrational excitations.

The emerging picture for the randomness-controlled Ioffe-Regel crossover which lies at the origin of the boson peak is schematically illustrated in Fig. 4.

\begin{figure}
\includegraphics[height=5.7cm,width=9.0cm]{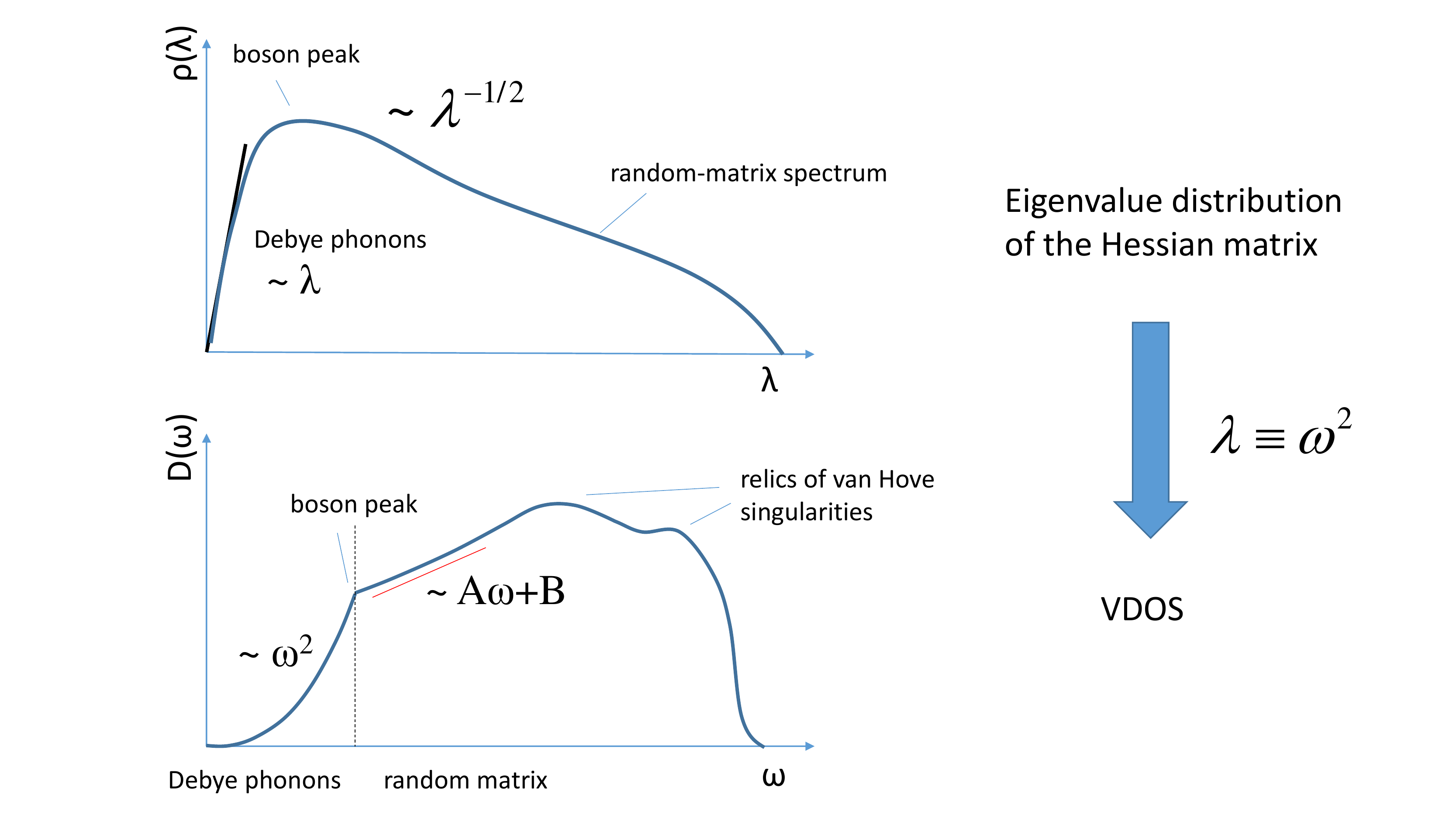}
\caption{Schematic illustration of the fundamental structure of the VDOS of disordered solids. The boson peak arises from the (Ioffe-Regel type) crossover from Debye phonons at low frequency into quasi-localized excitations the distribution of which is described by random-matrix statistics of the eigenvalues of the Hessian matrix~\cite{Baggioli_random}. }
\end{figure}

\section{The relation between relaxation time and vibration frequency in glassy systems}
\subsection{Relation between eigenfrequency and relaxation time}
A simple analytical relation between relaxation time and vibration eigenfrequency valid in an arbitrary solid with damping has been derived by Tobolsky~\cite{Tobolsky,Tobolsky_note}. 
The starting point is the usual Newton's equation for the dynamics of an atomic lattice, in presence of Markovian damping:
\begin{equation}
m\ddot{r}_{j}+\nu\dot{r}_{j}+\sum_{i}H_{ij}r_{i}=0
\end{equation}
where for simplicity we restrict ourselves to a 1d lattice, but the final result holds for arbitrary dimensions~\cite{Tobolsky_note}. In the above equation, $\nu$ is a viscous damping coefficient, $H_{ij}$ is the Hessian or dynamical matrix defined as usual (here for a 1d lattice), and $r_{j}(t)$ is the displacement from equilibrium of atom $j$. 
At the atomic level there is of course no frictional damping term, this is an emergent feature that arises from the dynamical coupling of atom $i$  with many atoms in the material, mediated by the long-range (anharmonic) part of the interaction potential. Such a frictional term can be recovered within particle-bath Hamiltonians~\cite{Zwanzig}, as will be done below in Sec. VI.C with more details, leading to a Generalized Langevin Equation which contains a frictional term. Under the assumption that the dynamic coupling between an atom and all the other atoms is chosen to be a constant, the Markovian damping of Eq. (13) above is retrieved. 

Upon rewriting the above equation in normal mode coordinates $\phi_p$ one obtains:
\begin{equation}
m\ddot{\phi}_{p}+\nu\dot{\phi}_{p}+k_{p}\phi_{p}=0
\end{equation}
with $p=1,2,...,N$, and $N$ the total number of atoms in the lattice, and where $k_{p}$ is the spring constant associated with eigenfrequency $\omega_p=\sqrt{k_{p}/m}$. With initial conditions $\phi_{p}(0)=0$ and $\dot{\phi}_{p}(0)=0$ the above equation is readily solved and the solution presents a combination of exponentials which contain the damping parameter $\nu$. Upon taking the strong damping limit, $\nu \gg \sqrt{4 m k_{p}}$, the following relation for the autocorrelation function of normal mode displacement is obtained:
\begin{equation}
\langle \phi_{p}(0) \phi_{p}(t) \rangle \approx \langle \phi_{p}^{2}(0) \rangle \exp(-t/\tau_{p})
\end{equation}
with the following fundamental relation between relaxation time and eigenfrequency valid for an arbitrary glassy (or simply anharmonic) solid in the strong damping limit:
\begin{equation}
\tau_{p}=\frac{\nu}{m\omega_{p}^{2}}
\end{equation}

It is important to note that, unlike $\tau_\alpha$ for example, $\tau_p$ is not a special time scale but simply a time scale associated with one of the $p=1...N$ normal modes (it would be $p=1...3N$ in 3d). 
The above relation is important as it establishes a crucial and direct connection between the relaxation spectrum or distribution of relaxation times $\tau_{n}$ in a glassy material and its VDOS. 
In spite of its importance, this relation has been substantially overlooked in previous work, aside from few lines of investigation. These are the fluctuating elasticity (FE) model introduced earlier, the Frenkel description of liquids~\cite{Trachenko1,Trachenko2,Trachenko2017}, which establishes a direct link between the frequency of $k$-gapped transverse modes and the Maxwellian relaxation time, and the nonaffine lattice dynamics (NALD) approach to glasses~\cite{Lemaitre,Milkus_PRE,Rodney}, which provides a direct link between VDOS and $\alpha$-relaxation~\cite{Cui_metal}.

Finally it is worth mentioning that in the study of the phonon Boltzmann transport equation using DFT, collective vibrational excitations, "relaxons", have been identified recently, which have a well defined relaxation time. 
This feature is important as it is directly related to heat-flux dissipation and relaxons can be used to provide an exact description of thermal transport in insulating crystals~\cite{Marzari_relaxon}.

\subsection{Relation between vibrational linewidth and relaxation time}
A natural question to ask is also about the relation between vibration lifetime (or linewidth, its inverse)
and relaxation time. 
\subsubsection{Supercooled liquids and melts}
In the hydrodynamic theory of liquids, such a relation emerges from the analysis of the coupled momentum transfer (Navier-Stokes) and heat transfer equations. Upon taking a double Fourier (in space) and Laplace (in time) transformation, the determinant of the hydrodynamic matrix of coefficients of the coupled Navier-Stokes and heat-transfer equations leads to factorization of longitudinal and transverse modes, with poles that contain imaginary parts that are related to heat diffusion and to sound attenuation.
For longitudinal modes, it is straightforward to show that the $\textbf{k}$-component of density relaxes as a function of time according to~\cite{Ruocco_RevModPhys}:
\begin{equation}
\rho_{\textbf{k}}(t) \sim \left[\exp(-D_{T}k^{2}t) +  \exp(-\Gamma k^{2} t) \right] \nonumber
\end{equation}
where we omitted prefactors, and $k$ denotes the wavevector. The first exponential is controlled by the thermal diffusivity $D_{T}$, whereas the second exponential is controlled by imaginary part of the acoustic dispersion, $\Gamma$, which is related to the viscosity of the material. $\Gamma$ coincides with the acoustic linewidth  which sets the lifetime of phonons, $\tau_{ph}$. Hence, in the limit of large thermal diffusivity $D_{L}$, the decay of density fluctuations is controlled by $\Gamma$ and the above equation establishes a simple relation between the vibrational lifetime $\tau_{ph}$ and the relaxation time $\tau$ of density fluctuations, $\tau_{ph} \sim \Gamma^{-1} \sim \tau$.
This relationship always holds in liquids as well as in amorphous solids in the hydrodynamic limit $k \rightarrow 0$. 

In supercooled liquids at finite wavelengths/frequencies, however, things are certainly not so simple~\cite{Ruocco_RevModPhys}. 
A simple, practical, but essentially empirical way of describing relaxation is by means of the viscoelastic model, where a memory function $M_{L} \sim \exp -(t /\tau(k))$ is introduced within the hydrodynamic treatment. Typically $\tau(k)$ has a peak that matches the first peak in the static structure factor $S(k)$, which denotes a slowing down of the dynamics due to cooperativity, a phenomenon known as de Gennes-narrowing. 
In general, $\Gamma(k)$ and $\tau(k)$ are not correlated for supercooled liquids, as demonstrated on the example of molten potassium in Refs.~\cite{Monaco2004,Cabrillo2002} where the viscoelastic relaxation time and the linewidth from either X-ray or neutron scattering are clearly uncorrelated as a function of $k$. 
In the case of liquid Argon, the linewidth displays a non-trivial dependence on $k$ as evidenced by Mode-Coupling theory and MD  simulations~\cite{Goetze_Zippelius}, and actually exhibits a minimum in correspondence of the first peak of $S(k)$ around $0.2$nm where the relaxation time $\tau(k)$ is expected to be largest. 
Recent MD studies have clarified that the de Gennes narrowing might actually be caused by geometric factors~\cite{Egami}, in particular by the long-range oscillations in $g(r)$, as demonstrated through the analysis of van Hove correlation function, rather than by collective dynamics. This is further supported by the fact that in real space $\tau(r)$ increases with $r$  linearly. These results call into question the identification of $\tau_{\alpha}$ with $\tau(k_{max})$ where $k_{max}$ is the position of the first peak in $S(k)$. 

\subsubsection{Glasses below $T_{g}$}
In glasses well below $T_g$, the situation is quite different. Here the linewidth is controlled by the disorder and is a purely harmonic phenomenon, at least at low enough temperatures. In this "harmonic disorder" regime, the linewidth goes as $\Gamma \sim \omega^{4}$~\cite{Schirmacher}, i.e. in agreement with the Rayleigh scattering law. Recently, numerical studies have shown that long-range power-law correlations in elastic constant may produce a logarithmic enhancement to this dependence~\cite{Gelin}. This effect has been theoretically predicted by extending FE theory to the case of long-range power-law correlations of the elastic constants~\cite{Cui2019}.
The relaxation process associated with the $\sim \omega^{4}$ harmonic disorder-induced scattering of vibrations is rather peculiar and has been elucidated using MD simulations of LJ glasses in ~\cite{RuoccoPRL2000}. In this process, the energy of an incoming plane wave gets redistributed to other eigenmodes of the glass without any viscous dissipation. 
Interestingly, using a MCT-type Generalized Langevin Equation approach to describe density correlations, Ruocco and co-workers arrive at the following relation between relaxation time of the harmonic wave-energy relaxation induced by disorder, $\tau$, and the linewidth, $\Gamma$:
\begin{equation}
\tau(k) \approx \frac{\Gamma(k)}{(v_{\infty}^{2} - v_{0}^{2})k^{2}}  \nonumber
\end{equation}
where $\Gamma$ is quadratic in $k$ and quartic (Rayleigh) in $\omega$, while $\tau$ is thus $k$-independent at low $k$, a result supported by recent simulations in the slow-dynamics regime~\cite{Sciortino2019}. In the above expression, $v_{\infty}$ and $v_{0}$ denote the speed of sound in the infinite and zero-frequency limit, respectively. 

Other relaxation channels involve quantum tunneling around $T \sim 1$K, and a relationship between ultrasonic attenuation and the time-correlation function dependent on the relaxation time has been presented within the framework of Two-Level State theory~\cite{Buchenau_TLS,Phillips_TLS}.
At higher $T$, in several systems also other relaxation channels become important which include thermally-activated hopping, an effect that can be measured with Brillouin scattering or ultrasound techniques, and controls the attenuation around $T=100$K~\cite{Vacher_damping}.
Anharmonic damping has been observed also in network glasses in the THz regime where it exhibits a viscoelastic-modified Akhiezer behaviour, with a Maxwell dependence on frequency:
\begin{equation}
\Gamma \sim \frac{\omega^{2} \tau}{1+\omega^{2}\tau^{2}} \nonumber
\end{equation}
with a single viscoelastic relaxation time $\tau$ and a viscoelastic crossover at $\omega \tau \sim 1$. At low frequency $\omega$, this recovers the hydrodynamic dependence $\Gamma \sim \tau \sim \eta$  where $\eta$ is the viscosity of the material.

\subsubsection{Organic liquids}
A comprehensive study of structural relaxation in the normal liquid phase of glycerol has been reported in Ref.~\cite{Giugni}, where different techniques (Brillouin spectra in the visible, UV, inelastic X-ray scattering and hypersonic measurements) were combined in an attempt to cover the broadest range of frequency.
Also in this case the frequency dependence of the longitudinal linewidth, $\Gamma(\omega)$, displays a viscoelastic crossover but this time it had to be fitted with a (more cooperative) Cole-Davidson model:
\begin{equation}
\Gamma(\omega)=\Gamma_{\infty}+(v_{\infty}^{2}-v_{0}^{2})\textrm{Re}\left(\frac{1-h(i \omega)}{i \omega}   \right)  \nonumber
\end{equation}
where 
\begin{equation}
h(i\omega)=\left(\frac{1}{1+i\omega \tau}\right)^{\beta}    \nonumber
\end{equation}
is the Cole-Davidson function, with $\tau$ the viscoelastic relaxation time and $\beta \leq 1$ the stretch parameter. The data taken over such a broad range of frequencies displayed a two-step viscoelastic relaxation. The above Cole-Davidson model had to be supplemented with an additional simple Maxwell ($\beta=1$) process which describes the first viscoelastic crossover at low frequency. This suggests that the first low-frequency process is a purely hydrodynamic one, similar to the one mentioned above for glasses below $T_{g}$ since in the hydrodynamic limit there ought to be no difference between liquid and glass. 
The second viscoelastic crossover in the normal liquid is probably related to caging and dynamical slow-down as described by MCT and related approaches.

\section{Linking vibrational and relaxation processes in metal alloys} 
In this section we will review the results obtained with theoretical approaches which have been proposed in an attempt to link the vibrational spectrum, i.e. the VDOS, with the relaxation features observed in material behaviour. We will focus on models which build a more direct connection between the VDOS and the stretched-exponential in $\alpha$ relaxation wing, since the latter process can be accessed experimentally in metal alloys~\cite{Wilde}.

\subsection{Mode-Coupling Theory}
Among the models which can describe the $\alpha$ relaxation in real materials, the more first principles, perhaps remains MCT although it does not provide a direct link between vibration and relaxation (though the VDOS can also be recovered from the MCT description of density fluctuations, as shown in ~\cite{Goetze_Mayr}, as mentioned above). In particular, MCT fittings have been proved able to describe the $\alpha$ peak and the 
$\alpha$ relaxation wing of organic molecular glass formers such as propylene carbonate in Ref.~\cite{Blochowitz}, with few fitting parameters. Fittings of the loss viscoelastic modulus of colloidal glasses in comparison with rheological measurements data have also been reported in recent work~\cite{Laurati}.

\subsection{The fluctuating elasticity (FE) model}
As mentioned in Sec. IIIA, the FE model assumes, as its starting point, quenched Gaussian disorder in the elastic shear modulus. Referring to the simplest version of the theory and focusing on longitudinal excitations, the dynamical susceptibility is found within this approach to be
\begin{equation}
\chi_{L}(k,z)=\frac{k^{2}}{-z^{2}+k^{2}c_{L}^{2}(z)}=k^{2} G_{L}(k,z)
\end{equation}
where, again, $z=\omega+i0^{+}$, while $c_{L}(z)$ is the complex longitudinal speed of sound, and $G_{L}(k,z)$ is the longitudinal disorder-averaged Green's function obtained from the FE approach as outlined in Sec. IIIA. The complex viscoelastic modulus is thus obtained as
\begin{equation}
E(z)=\rho c_{L}(z)^{2}=E'(\omega) - iE''(\omega)
\end{equation}
This is actually a generalization of the FE to fluctuating viscoelastic moduli, also called Heterogeneous Viscoelasticity~\cite{Schirmacher_viscoelastic}. 

This theory provides a direct link between the VDOS and the $\alpha$-relaxation process because the Green's function $G(z)$ is the same input used to compute the VDOS as mentioned in Sec. IIIA as well as to compute the viscoelastic loss modulus $E''(\omega)$ where the $\alpha$-relaxation process is visible as a wing to the high-frequency side of the loss peak.

In its more sophisticated version where the Green's function is evaluated using Coherent Potential Approximation (CPA), the theory yields the following description of the $\alpha$-relaxation wing in the loss modulus, shown in Fig. 5.

In this comparison there are a few fitting or adjustable parameters, which coincide with the parameters that define the Gaussian function (see inset of Fig. 5) used to describe the spatially-fluctuating distribution of the high-frequency shear modulus, in turn related to the spatial distribution of energy barriers (dynamical heterogeneity) through the shoving model of the glass transition~\cite{Dyre_1998,Dyre_review}. 
Thus the same "disorder parameters" which define the Gaussian distribution for the fluctuating elasticity control both the VDOS of the system and the $\alpha$-relaxation wing visible in Fig. 5. Upon Fourier-transforming the latter into the time-domain, a function of time similar to a stretched-exponential can be obtained. Also, the same model also predicts a boson peak in the specific heat, according to Eq. 2 ~\cite{Schirmacher_viscoelastic}. 

\begin{figure}
\includegraphics[height=5.7cm,width=8.0cm]{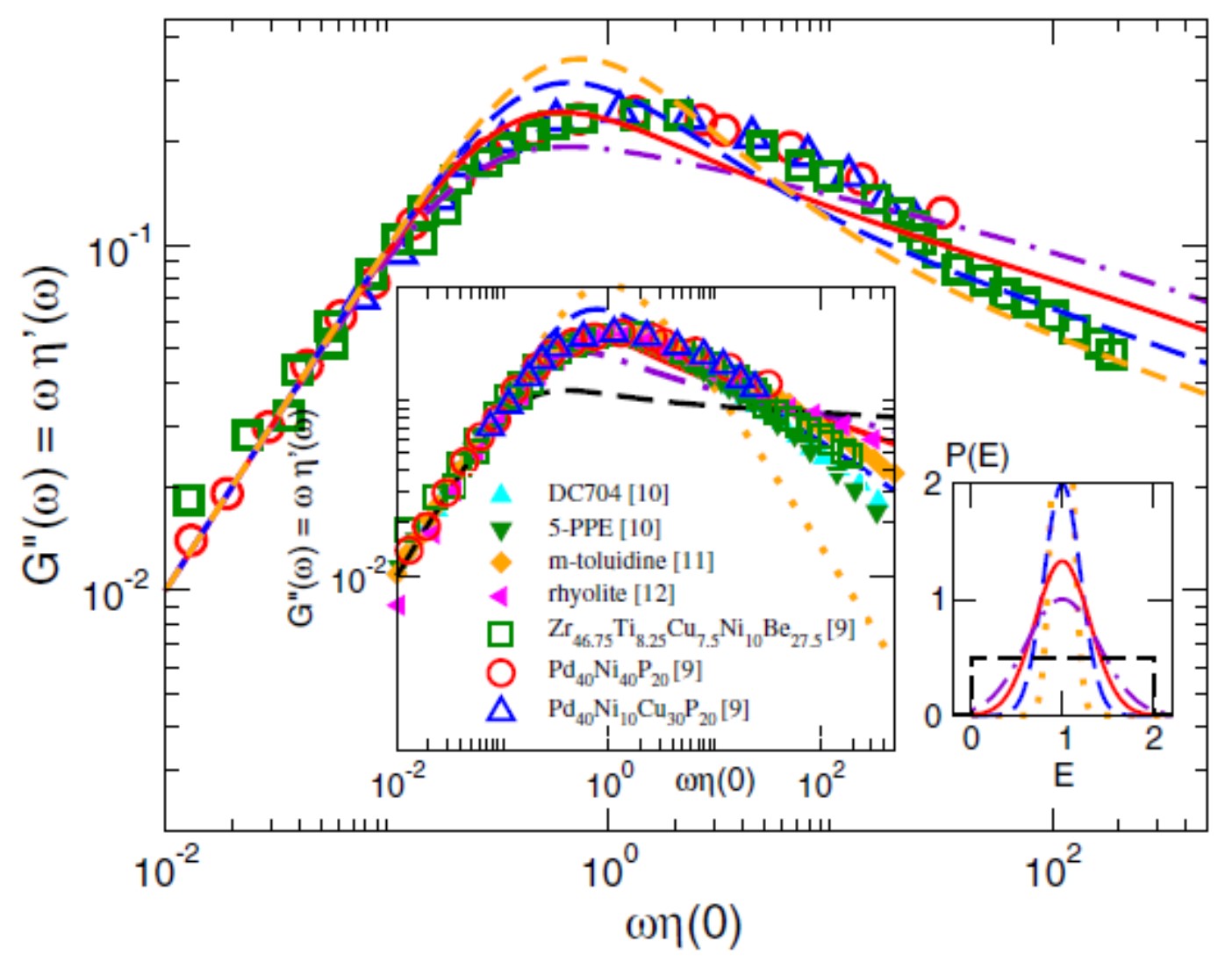}
\caption{Comparison between the Heterogeneous Viscoelasticity theory description (based on the FE model) and various sets of experimental data (including metallic glasses) for the shear viscoelastic loss modulus $G''(\omega)$ as a function of mechanical oscillation frequency $\omega$. Adapted with permission from Ref.~\cite{Schirmacher_viscoelastic}, Copyrights American Physical Society (2019).}
\end{figure}

Finally, at high frequencies this theory recovers the FE model, which indicates that the FE model works within the frame of affine elasticity. It is known, indeed, that the frequency-dependent elastic modulus of an amorphous material reaches a high-frequency plateau that coincides with the affine elasticity approximation~\cite{Milkus_PRE}. Implementation of nonaffine elasticity within the FE and the Heterogeneous Viscoelasticity frameworks remains a task for ongoing and future work. 

\subsection{Linking vibration and relaxation in metal alloys with the NALD approach}
\subsubsection{The viscoelastic NALD approach}
The starting point of the Nonaffine Lattice Dynamics (NALD) approach is to consider a tagged atom $i$ and to model it as interacting with its nearest-neighbours (through the Hessian, in the same way as in Eq. (13) above) and then also with all other particles (atoms) in the material to which it is coupled dynamically via a bilinear term in the Hamiltonian.
Effectively the tagged atom behaves like a tagged particle that lives in a local environment (a force-field) given by its nearest-neighbours, and in addition its dynamics is coupled to all other atoms in the system (the bath oscillators), which is an effective way of modelling long-range many-body interactions.  This class of Hamiltonians are known as particle-bath or Caldeira-Leggett Hamiltonians and their classical (non-quantum) formulation is due to Zwanzig~\cite{Zwanzig}. 

The Hamiltonian of the tagged atom coupled to its nearest-neighbours as well as all the many other atoms in the material (treated as harmonic oscillators) is given by~\cite{Zwanzig}
\begin{equation}
H=\frac{P^2}{2m}+V(Q)+\frac{1}{2}\sum_{\alpha=1}^N\left[\frac{p_{\alpha}^2}{m_{\alpha}}+m_{\alpha}\omega_{\alpha}^2\left(X_{\alpha}
-\frac{F_{\alpha}(Q)}{m_{\alpha}\omega_{\alpha}^2}\right)^{2}\right]
\end{equation}
where the first two terms are the Hamiltonian of tagged particle with (effective) mass $m$ and coordinate $Q$, while $\frac{1}{2}\sum_{\alpha=1}^N(\frac{p_{\alpha}^2}{m_{\alpha}}+m_{\alpha}\omega_{\alpha}^2 X_{\alpha}^2)$ is the Hamiltonian of the bath of harmonic oscillators coupled to the tagged particle with linear coupling function $F_{\alpha}(Q)=c_{\alpha}Q$ where $c_{\alpha}$ are the coupling strength coefficients which are different for all the different atoms the tagged atom is interacting with (e.g. $c_{\alpha}$ is expected to be large for nearby atoms and small for atoms far away in the material).

This configuration gives rise to a second-order inhomogeneous differential equation for the position of the $\alpha$-th oscillator of the bath, whose solution leads to the following GLE~\cite{Zwanzig}:
\begin{equation}
m\ddot{Q}=-V'(Q)-\int_{-\infty}^t\frac{\nu(t-t')}{m}\frac{dQ}{dt'}dt' + F_{p}(t).
\end{equation}
where $F_{p}(t)$ is the thermal noise, with the usual fluctuation-dissipation statistics (which, however, acquires new terms when the system is subjected to an external oscillatory field~\cite{Cui_FDT}).

As is standard for normal mode analysis, one can introduce the rescaled tagged-particle displacement $q=Q\sqrt{m}$ in the Hamiltonian, such that the resulting equation of motion, using mass-rescaled coordinates, becomes
\begin{equation}
\ddot{q}=-V'(q)-\int_{-\infty}^t \nu(t-t')\frac{dq}{dt'}dt' + F_{p}(t).
\end{equation}
The noise term can be shown to drop out upon taking an ensemble average, or equivalently, upon averaging over cycles.

As shown in Ref. \cite{Zwanzig},
the friction coefficient $\nu$ arises from the long-range coupling between atoms in the particle-bath model, which effectively takes care of long-range and many-body anharmonic tails of interatomic interaction. The friction, in general, arises due to the coarse-graining of a number of degrees of freedom (basically all the coordinates and momenta of the bath particles, as well as the momentum of the tagged particle). The friction is typically non-Markovian (i.e. time or history-dependent) and given by a memory kernel, in the most general case, as is also the case in the MCT equations.

Upon applying a deformation described by the strain tensor $\underline{\underline{\eta}}$, the nonaffine dynamics of a tagged particle $i$ interacting with other atoms satisfies the following equation for the displacement $\{\underline{x}_i(t)=\underline{\mathring{q}}_i(t)-\underline{\mathring{q}}_i\}$ around a known rest frame $\underline{\mathring{q}}_i$ (see Ref.~\cite{Cui_metal} for details of derivation):
\begin{equation}
\ddot{\underline{x}}_i +\int_{-\infty}^{t}\nu(t-t')\dot{\underline{x}}_i dt'+\underline{\underline{H}}_{ij}\underline{x}_j=\underline{\Xi}_{i,xx}\eta_{xx},
\end{equation}
which can be solved by performing Fourier transformation followed by normal mode decomposition that decomposes the 3N-vector $\tilde{\underline{x}}$, that contains positions of all atoms, into normal modes $\tilde{\underline{x}}=\hat{\tilde{x}}_p(\omega)\underline{\phi}^p$ ($p$ is the index labeling normal modes).  Note that we specialize on time-dependent uniaxial strain $\eta_{xx}(t)$. For this case, the vector $\underline{\Xi}_{i,xx}$ represents the force per unit strain acting on atom $i$ due to the motion of its nearest-neighbors which are moving towards their respective affine positions (see e.g.Refs.~\cite{Lemaitre,Scossa-Romano} for the analytical expression and derivation) and in the case of metal alloys it also includes electronic effects empirically via the Embedded Atom Model (EAM) potential~\cite{EAM}.\\

Within NALD, the stress can be derived upon taking the total derivative of the internal energy associated with the above equation of motion, by imposing mechanical equilibrium at all steps in the deformation. Upon further performing a normal mode analysis and upon computing the stress, one arrives at the following expressions for the stress (in the case of longitudinal deformation along the $x$-axis):
\begin{align}
\tilde{\sigma}_{xx}(\omega)&=E_A\tilde{\eta}(\omega)-\frac{1}{V}\sum_p\hat{\Xi}_{p,xx}\hat{\tilde{x}}_p(\omega) \notag\\
&=E_A\tilde{\eta}(\omega)+\frac{1}{V}\sum_p\frac{\hat{\Xi}_{p,xx}\hat{\Xi}_{p,xx}}{\omega^2-\omega_p^2
-i\tilde{\nu}(\omega)\omega}\tilde{\eta}(\omega)\notag\\
&=E_{xxxx}(\omega)\tilde{\eta}(\omega).
\end{align}
from which the viscoelastic Young modulus is readily obtained:
\begin{equation}
E^*(\omega)=E_A-3\rho\int_0^{\omega_{D}}\frac{D(\omega_p)\Gamma(\omega_p)}{\omega_p^2-\omega^2+i\tilde{\nu}(\omega)\omega}d\omega_p
\end{equation}
where we have dropped the Cartesian indices for convenience, since we are specializing on uni-axial extension, and $\rho=N/V$ s the atomic density of the solid.\\

Here we have used $\omega_{p}$ to denote the eigenfrequency to avoid ambiguity with $\omega$ which denotes the frequency of the applied mechanical oscillation. 

Using the EAM method to describe interatomic interactions in metal alloys, it has been possible to evaluate the above expression for the case of binary $\textrm{Cu}\textrm{Zr}$ alloys. The comparison (taken from Ref.~\cite{Cui_metal}) is shown in Fig. 6 below. 

\begin{figure}
\includegraphics[height=5.7cm,width=8.0cm]{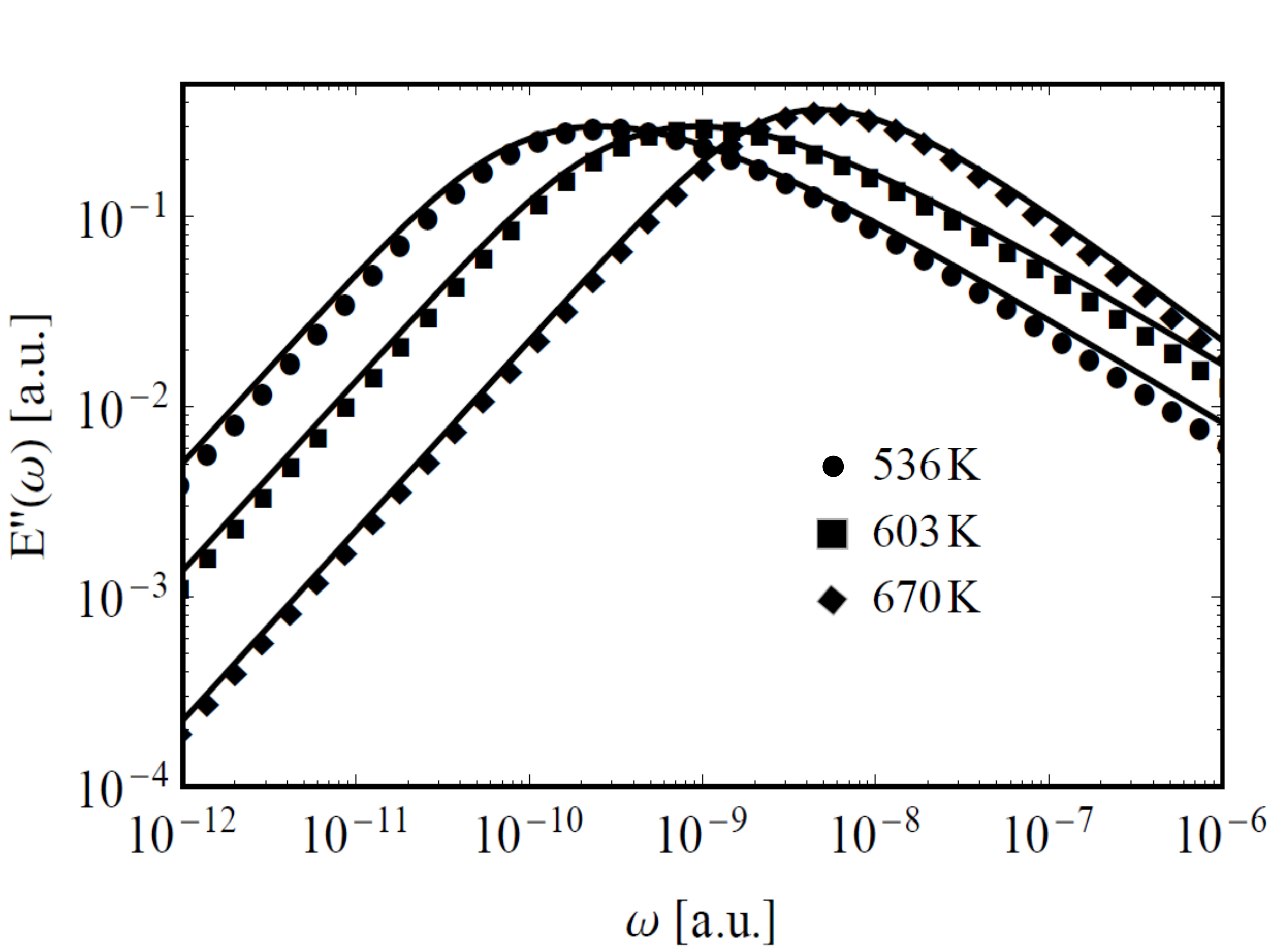}
\caption{Comparison between the NALD theoretical description and experimental data for the Young viscoelastic loss modulus $G''(\omega)$ of $\textrm{Cu}_{50}\textrm{Zr}_{50}$ alloys as a function of mechanical oscillation frequency $\omega$. The temperature range is from slightly above to slightly below the glass transition temperature. Adapted with permission from Ref.~\cite{Cui_metal}, Copyrights American Physical Society (2019).}
\end{figure}

\subsubsection{Memory function to bridge from microscopic to macroscopic relaxation}
In the above comparison few fitting parameters are used in the modelling of the memory kernel, which is taken to be a stretched-exponential function. The latter choice is motivated by the fact that, as within MCT schemes, the memory kernel dictates the shape of the intermediate scattering function~\cite{Baity}. A beautiful relation between the memory kernel and the intermediate scattering function was derived from first-principles by Sjoegren and Sjoelander~\cite{Sjoegren}
\begin{equation}
\nu(t)=\frac{\rho k_{B}T}{6\pi^2 m}\int_{0}^{\infty}dk k^{4} F_{s}(k,t)[c(k)]^{2} F(k,t)
\end{equation}
where $c(k)$ is the direct correlation function of liquid-state theory, $F_{s}(k,t)$ is the self-part of the intermediate scattering function and $F(k,t)$ is the intermediate scattering function~\cite{Sjoegren}. All of these quantities are functions of the wave-vector $k$ and the integral over $k$ leaves a time-dependence of
$\nu(t)$ which is exclusively given by the product $F_{s}(k,t)S(k,t)$. Similar relations between time-dependent friction and intermediate scattering function have been derived within the frame of the Self-Consistent Generalized Langevin Equation (SCGLE) theory by Medina-Noyola and co-workers and applied to colloidal systems~\cite{Medina1,Medina2} and monoatomic liquids~\cite{Medina_atomic}. 

From approximate fittings to the solution to MCT equations, and also from scattering experiments and simulations, we know that in supercooled liquids $F(k,t)\sim \exp(-t/\tau)^\beta$, with values of the stretching exponent that normally lie in the range $\beta=0.6-0.8$~\cite{Ruta}.
In turn, this argument gives $\nu(t)\sim \exp[-(t/\tau)^b]$, with stretching exponent $b$ for the memory kernel.
For $\textrm{Cu}_{50}\textrm{Zr}_{50}$ near $T_{g}$ in the comparison in Fig. 6, it was found $b=0.58-0.75$ for the memory kernel $\nu(t)$, corresponding to $\beta=0.76-0.85$ in $F(k,t)\sim \exp(-t/\tau)^\beta$ using Eq. (25), which is in good agreement with experimental determinations via X-ray photon-correlation spectroscopy (XPCS) for supercooled metallic melts~\cite{Ruta}, where $\beta\approx 0.8$ in the supercooled regime near the glass transition temperature. Also,$\beta = 0.76$ was found in the normal liquid state with quasielastic scattering~\cite{Meyer}. 
From direct stretched-exponential fittings of $E''(\omega)$ in DMA data, however, typically one finds stretch exponents which are systematically lower, around $\beta \approx 0.4 - 0.5$~\cite{Carter,Wang_JCP,Pineda}. 

A possible resolution to this discrepancy is indicated by Eq. (25) above and by the model fitting of Ref. ~\cite{Cui_metal}, where the stretch exponent of the memory kernel for the fitting of $E''(\omega)$ (which is closer to the stretch exponent that one gets from a direct stretched-exponential fitting of macroscopic data) is basically the square of the stretch exponent of $F(k,t)$.

The advantage of this approach is the transparent and direct link between the VDOS and the macroscopic material response, e.g. the loss modulus, hence the relaxation behaviour. Upon taking the memory kernel as a stretched-exponential function, the only fitting parameters are the prefactor, the characteristic relaxation time and the stretching-exponent $b$. As mentioned already, however, $b$ is found to agree with experimental measurements~\cite{Ruta,Ruta_review}  in the supercooled regime, which effectively reduces the number of fitting parameters to two non-trivial adjustable parameters. 

It will be shown later on in the context of dielectric relaxation, that the $\alpha$ wing and hence the stretched-exponential in the material response are recovered by the GLE model even for constant (Markovian) friction, and a strong correlation has been shown (on toy models) between the boson peak and the extent of the $\alpha$ wing and the stretching exponent. 

\subsubsection{Separation of timescales in XPCS and DMA}
We have noted that the characteristic time-scale probed in DMA is significantly larger than the time-scale measured in XPCS and quasielastic scattering experiments. The only systematic study available on the same system, $\textrm{Mg}_{65}\textrm{Cu}_{25}\textrm{Y}_{10}$, has evidenced that the relaxation timescales probed by DMA are on the order of $\tau = 10^{-1} - 10^{0}$s, while the timescales probed in XPCS are in the range $\tau = 10^{2} - 10^{4}$. This wide separation of timescales (at least two orders of magnitude in $\tau$) may also suggest that, in fact, widely different relaxation processes altogether might be at play. This is an important point which  deserves more extensive and systematic studies across multiple time and length-scales possibly combining numerical simulations, experiments and theoretical approaches.

\subsubsection{Modelling of $\alpha$ and $\beta$ relaxation}
Finally, the NALD approach has been used also to describe both $\alpha$  and $\beta$ relaxation in metal alloys. Also in this case, the VDOS  extracted from experiments (or alternatively, from MD simulations) was implemented in Eq. (24), for the paradigmatic case of a ternary alloy, $\textrm{La}_{60}\textrm{Ni}_{15}\textrm{Al}_{25}$.
In this system there is one atomic component, $\textrm{La}$, which is much heavier than the other two. This fact introduces a separation of time scales in the dynamics that shows up in distinct peak contributions in the VDOS. In order to reproduce both $\alpha$ and $\beta$ relaxation processes in the fitting of the viscoelastic loss modulus $G''(\omega)$ it is necessary to implement two terms in the memory kernel, $\nu(t)=\nu_{1}(t)+\nu_{2}(t)$, both modelled as stretched-exponentials, with different characteristic times. It was found that the $\beta$ relaxation process receives a dominant contribution from the dynamics of the $\textrm{Al}$ atoms, which are the lightest in terms of atomic mass. This finding is consistent with the picture of activating units of atoms which are able to perform thermally-activated jumps outside the cage, where, in a system with wide separation of atomic masses, the lightest atoms are going to be the majority components of the activating units.
Overall, these findings point toward the existence of degrees of freedom with well separated time-scales as the possible origin of $\beta$-relaxation processes and this idea should be explored further in future studies possibly in combination with the so-called coupling model~\cite{NGai1998}.

\section{Vibration and relaxation in metallic glasses: power laws}
The VDOS of amorphous metal alloys (metallic glasses) can be accessed via inelastic neutron scattering (INS) as well as from MD  simulations. Among the earliest measurements are those carried out by Suck and co-workers~\cite{Suck1980,Suck1981}. Interestingly, these results (later confirmed by many other investigators) evidenced the existence of a power-law regime in the VDOS in an energy window near the boson peak, and in particular between 4.7 meV  and 8 meV, where the VDOS exhibits the following power-law behaviour in glassy CuZr alloys:
\begin{equation}
D(\omega) \sim \omega ^{4/3}.
\end{equation}
In spite of the large amount of experimental and numerical studies that followed (and substantially confirmed) the data of Suck et al., the emergence of this power-law behaviour has remained largely unexplained to date. 

It is tempting to interpret the power-law in the VDOS of CuZr alloys measured experimentally ~\cite{Suck1980} in light of the fracton model of Alexander and Orbach~\cite{Alexander-Orbach} (reviewed in Sec. III.E) and of recent structural results which unveiled fractal patterns in the atomic packing and density correlations of metallic glasses. 
In particular, both neutron and X-ray diffraction results~\cite{Ma2008} as well as numerical MD simulations~\cite{Yang2017}, suggest the existence of fractal correlations with a fractal exponent $\bar{d}=2.3$ in the
relation between the number and size of activating units. The activating units are defined as clusters of mobile atoms which display thermally activated jumps beyond nearest-neighbour distances. Similar clusters have been detected also in entropically-stabilized BCC (Wigner) crystals of charged colloidal particles~\cite{Sprakel}. 

Hence, we can set $\bar{d}=2.3$ for the fractal dimension of the activating-clusters network, motivated by experiments and simulations~\cite{Ma2008,Yang2017}. Furthermore, we can set $\bar{\delta}\approx 0$ which corresponds to normal diffusion, because the atoms in the activating units effectively diffuse on the length scale of the cluster due to the thermally activated jumps. Hence, we obtain $[2\bar{d}/(2+\bar{\delta})]-1 \approx 1.3$ which is very close to the exponent $4/3=1.33$ measured by Suck et al. ~\cite{Suck1980}. 

This new interpretation of the $4/3$ law in the VDOS of amorphous metal alloys based on the fracton model, proposed here, may need to be further supported in future work. This could be done, for example, by comparing the characteristic energy of the activating units which form the fractal network with the energy window where the $4/3$ law is observed in the VDOS for specific systems. 

If this picture will be confirmed, it may indicate that within this energy interval the dominant contribution to the VDOS comes from the fractal network of activating units, within which vibrational excitations propagate in a diffusive-like manner. This may then imply that the boson peak in metallic glasses coincides with a Ioffe-Regel crossover~\cite{Aharony} from ballistic Debye phonons into quasi-localized excitations that move diffusively along the fractal network of activating units.

Also, the identification of fracton excitations may be important to provide a more solid theoretical framework for the description of $\beta$ relaxation in amorphous metal alloys, which is due to stringy and cluster-like motions of mobile atoms~\cite{Samwer}. This could be a possible route towards linking vibrational (fracton) properties and $\beta$ relaxation in metal alloys.

Finally, power-laws in amorphous metal alloys have been observed also in avalanches, i.e. intermittent stress drop events associated with plastic rearrangements under external strain~\cite{Zapperi,Robbins}. Numerous studies have shown that avalanches in amorphous solids follow power-law statistics $P(x)\sim x^{-s}$ where $x$ is a variable that characterizes the size of the avalanche, and recently these phenomena have been observed experimentally in metallic glasses~\cite{Lagogianni}. 
A further step could be to link mean-field analytical models of avalanche statistics~\cite{Dahmen} with fracton excitations in an attempt to produce a unifying picture of power-laws emerging in amorphous metals.

\section{Relaxation and vibration in molecular glass formers}
Dynamical relaxation in the liquid and glass state has been studied intensely by means of Broadband Dielectric Spectroscopy (BDS) over the last decades, using small organic molecules (glycerol, propylene carbonate, etc) as model systems. BDS allows one to access the loss dielectric modulus $\epsilon''(\omega)$ over many orders of magnitude in electric field frequency $\omega$. Such a breadth of frequency range is only possible with dielectric spectroscopy techniques, and it has provided deep insights into the various relaxation phenomena occurring in supercooled liquids as temperature evolves upon going from the liquid to the glassy state~\cite{PLunk2000,Blochowitz_review} as well as in disordered solids, such as  orientationally-disordered (plasic) crystalline phases~\cite{Brand2002}, in general. 

In spite of this large body of experimental work, theoretical work has been comparatively much more limited, partly due to the difficulty of modelling such strongly interacting molecular systems. 
Modelling has been mostly limited to empirical models such as Havriliak-Negami or Cole-Cole relations, which consist of empirical ''decorations" of the classic Debye relaxation model.
The Debye relaxation assumes that molecules can be described as independently reorienting dipoles in the alternating electric field, and solves the Smoluchowski diffusion equation for the angular coordinates of the orienting dipole. The time-dependent solution, as for any diffusive process, is of course just a simple decaying exponential in time. Therefore, the time-dependent dielectric response goes like $\epsilon \sim \exp(-t/\tau)$ with a single characteristic relaxation time $\tau$ which is related to the specifics of the molecular dipole. 
Upon Fourier transforming, the dielectric complex modulus is 
\begin{equation}
\epsilon^{*}(\omega)=\epsilon_{\infty} + \frac{\epsilon_{S}-\epsilon_{\infty}}{1+i\omega \tau}
\end{equation}
where $\epsilon_{S}$ is the static electric constant, while $\epsilon_{\infty}$ is the infinite frequency value. The imaginary part of the above expression gives the loss modulus $\epsilon''(\omega)$ which features a symmetric absorption peak centered on a characteristic frequency $\omega \sim 1/\tau$. 

Since the Debye relaxation assumes independently reorienting dipoles, it neglects the cooperativeness of collective motions which is of course crucial for the description of supercooled liquids and glasses in the $\alpha$ relaxation regime, and which shows up as an asymmetric wing to the absorption peak, as already discussed above. The Havriliak-Negami, Cole-Cole, Cole-Davidson and other similar models empirically modify the Eq. (27) by elevating the denominator and the term $(i\omega \tau)$ within it, to power-law exponents which  are then fitted to the experimental data. In general, all these models can provide fittings of data but do not provide insights into the physics of relaxation.

For $\beta$ relaxation, typically a Cole-Cole empirical function is used for the modelling of this relaxation process in BDS experiments. A more physical model is the one proposed by Colmenero and co-workers which is based on modelling the secondary $\beta$ relaxation process by taking a Gaussian distribution of activation energies together with the Arrhenius thermal activation law~\cite{Colmenero}.

Other than empirical models, more physical and first-principles descriptions are very rare, and basically reduce to two approaches, that have already been mentioned above: the MCT approach, and the NALD-GLE approach. We will briefly review some results from these two approaches.

\subsection{MCT description of dielectric relaxation of glasses}
Within the schematic MCT framework (i.e.  the one where the wavevector in $F(q,t)$ is fixed to be equal to the cage size, corresponding to the first peak of the static structure factor $S(q)$), $\alpha$ relaxation and stretched-exponential behaviour can be reproduced upon suitably accounting for thermally-activated intermittent jumps using ad hoc memory kernels. The main shortcoming of the Mori-Zwanzig projection operator formalism on which MCT relies is that it is not trivial to accommodate thermally-activated processes within projection operators. However, certain forms of the correlators which make up the memory kernel of the MCT equation of motion have been shown to be able to reproduce, to some extent, the intermittent jumps~\cite{Blochowitz}. 

Using two distinct contributions to the memory kernel, one stemming from collective density fluctuations and the other from the correlations between a probe-particle and the rest of the system, Voigtmann and co-workers ~\cite{Blochowitz} have managed to provide an excellent description of both $\alpha$ and secondary $\beta$ relaxation in propylene carbonate, as shown in Fig. 7.

\begin{figure}
\includegraphics[height=5.7cm,width=8.0cm]{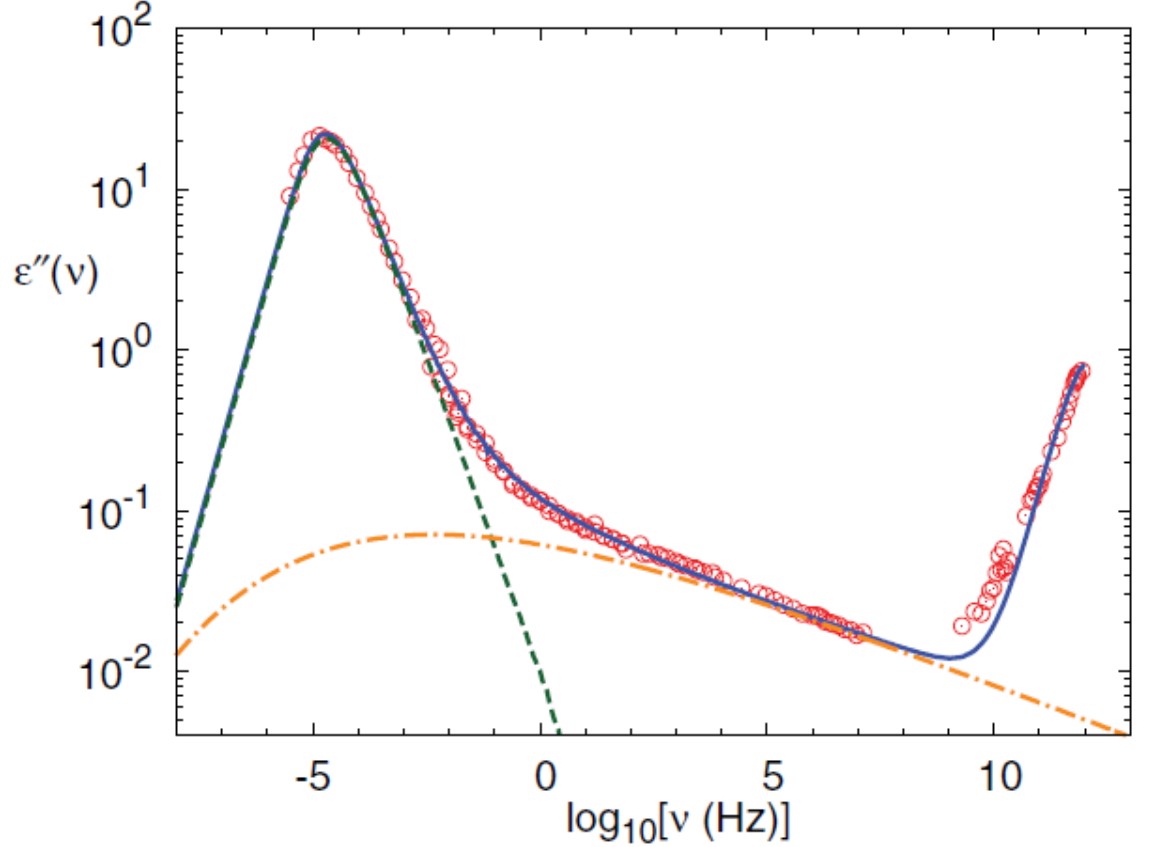}
\caption{Comparison between MCT theoretical calculations (solid line) of dielectric loss $\epsilon''(\nu)$ as a function of electric field frequency $\nu$ at $T=157 K$ and experimental BDS data for propylene carbonate (symbols) The dashed line represents the calculation with just one contribution to the memory kernel, namely the collective density fluctuations, which are shown to capture the $\alpha$ relaxation peak. The dashed-dotted line represents the calculation with just the other contribution to the memory kernel, which is due to coupling between probe-molecule and collective fluctuations, shown to capture the secondary $\beta$ relaxation process. Adapted with permission from Ref.~\cite{Blochowitz}, Copyrights American Physical Society (2019).}
\end{figure}

Clearly, future work in this area will be addressed towards strengthening the connection between the analytical forms of the contributions to the memory kernel and the underlying microscopic dynamical processes. This approach also has the potential of providing a way of connecting the material response, in terms of $\epsilon''$ with the VDOS  by implementing the same physics as encoded in the memory functions inside the calculation scheme for the VDOS based on MCT description of density fluctuations~\cite{Goetze_Mayr}. 

\subsection{GLE-NALD description of dielectric relaxation of glasses}
An approach similar to the one described in Sec. VI.C for the viscoelastic loss modulus can be also applied to describe the dielectric loss modulus of supercooled liquids and glasses. 

Within this picture, positive and negative partial charges in the polarized molecules form a network or lattice, similar to an ionic lattice, as schematically depicted in Fig. 8. 

\begin{figure}
\includegraphics[height=5.0cm,width=8.2cm]{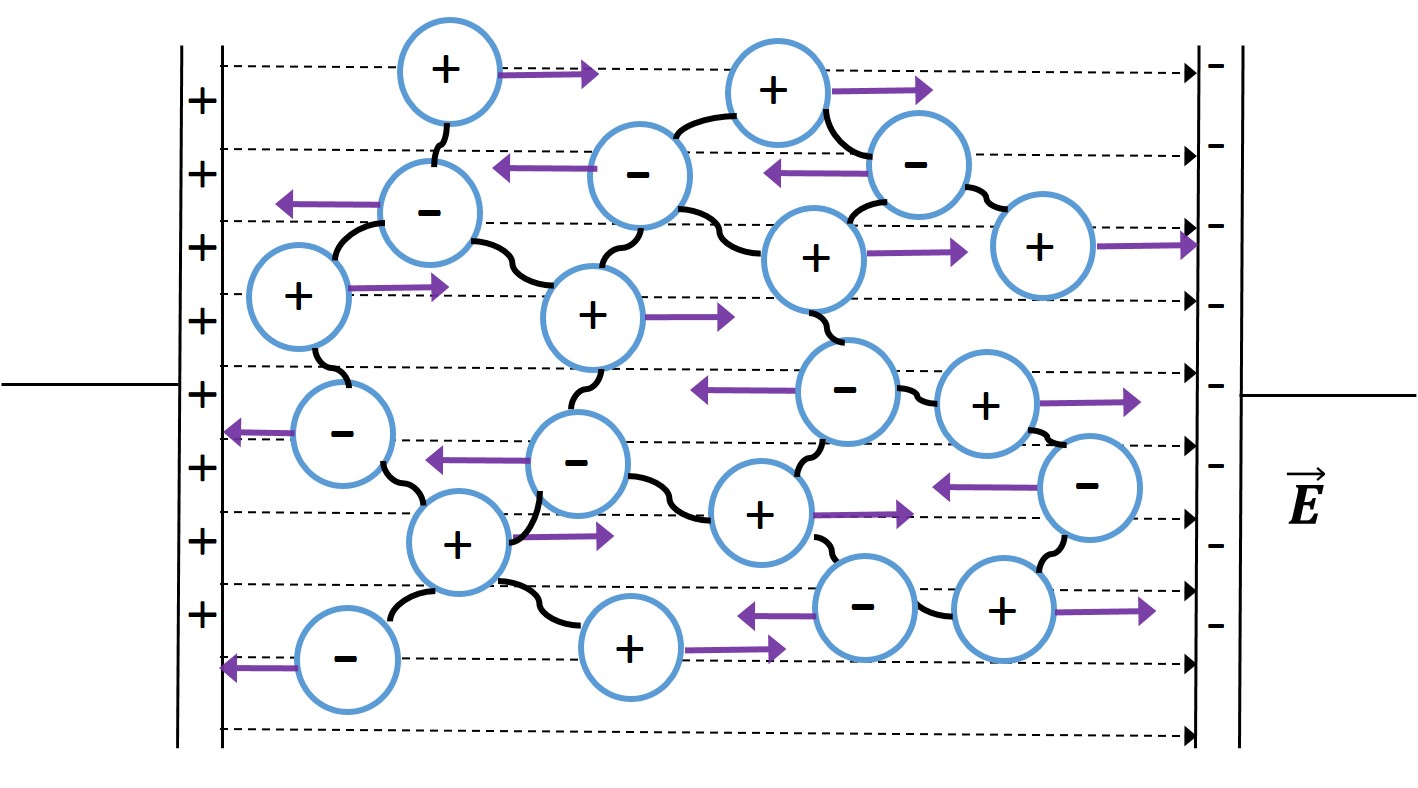}
\caption{Schematic illustration of the modelling strategy used within the NALD-GLE approach to describe the motions of partial negative and positive charges of polarized molecules in an electric field in dielectric spectroscopy experiments. Each positive or negative partial charge interacts with its neighbours with an effective interaction. Furthermore, it is also dynamically coupled to all other particles (via long-range interactions) which is taken into account via the bi-linear particle-bath oscillator coupling, the last term on the r.h.s. in Eq. (19). }
\end{figure}

Each of these partial charges, that we may call "particles", interacts with some neighbours via some effective interactions which define an effective spring constant. Furthermore, long range (Coulombic) interactions establish dynamical couplings between each of these partial charges and all the others in the system, in the spirit of particle-bath dynamical coupling. Hence, one can use the Hamiltonian Eq. (19) where there is a term describing the local interaction with neighbours and a  term (the last one on the r.h.s. of Eq. (19)) which instead implements the long-range coupling between the tagged particle and all the other particles in the system.

Also in this case, upon eliminating fast variables (momenta)  from the Hamiltonian Eq. (19), one arrives at the GLE, Eq. (21). Now, in the case of dielectric relaxation one does not have to worry about nonaffine displacements, since due to the electric field being uniform in the $x$ direction and only the $x$-component of the response matters, the microscopic displacements are all affine because the nonaffine components vanish by symmetry. Since in this case the external forcing term is not the applied strain, but the applied (oscillatory) electric field, the term on the r.h.s. side will be different from the one in Eq.(22) and will be given, instead, by the electric force acting on the tagged particle:
\begin{equation}
m\ddot{ \uline{r}}_{i}+\int_{-\infty}^t \nu_0e^{-[(t-t')/\tau]^b}\dot{\uline{r}}_{i}dt'+\uline{\uline{H}}_{ij}\uline{r}_{j}=q_{e}\uline{E}.
\end{equation}
where we have already explicity implemented a stretched-exponential expression for the memory kernel.

Upon performing the usual normal mode decomposition and upon Fourier transforming, we finally obtain the following expression for the complex dielectric modulus:
\begin{equation}
\epsilon^*(\omega)=1-A\int_{0}^{\omega_{D}}\frac{D(\omega_{p})}{\omega^2-i( \tilde{\nu}(\omega)/m)\omega-C^2\omega_p^2}
d\omega_{p}
\end{equation}
where $\omega_{p}$ denotes the normal mode frequency, whereas $\omega$ denotes the frequency of the external electric field. In the above equation, $A$ is a rescaling constant, since experimental data are normally given in arbitrary units, whereas $C=\sqrt{k/m}$ is related to the average effective spring constant $k$ of nearest-neighbour interactions and is treated as a constant parameter, while $\nu(\omega)$ is the Fourier transform of the memory kernel $\nu(t)=\nu_0 e^{-[(t-t')/\tau]^b}$ in Eq. (28). 

Upon taking the imaginary part of Eq. (29), the dielectric loss modulus $\epsilon''$ is readily obtained. This has been used in Ref.~\cite{Cui_dielectric} to fit experimental BDS data of glycerol obtained by the Augsburg group of Loidl, Lunkenheimer and co-workers~\cite{Loidl}. The comparison is shown in Fig. 9.

\begin{figure}
\includegraphics[height=5.7cm,width=8.0cm]{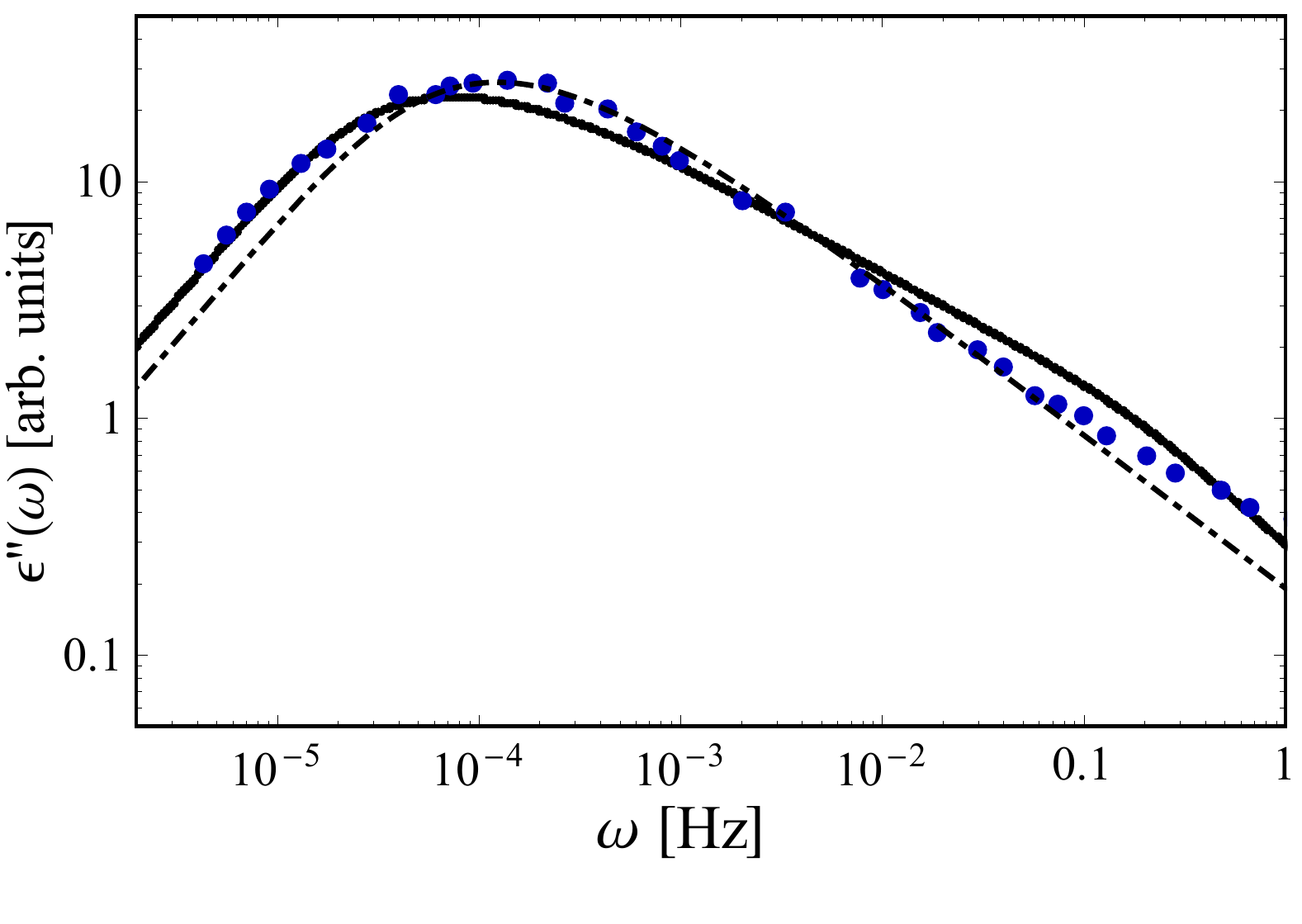}
\caption{Comparison between the NALD-GLE theoretical calculations (solid line) of dielectric loss $\epsilon''(\omega)$ as a function of electric field frequency $\omega$ at $T=184 K$ and experimental BDS data for glycerol (symbols) measured in Ref.~\cite{Loidl}. The dashed line represents an empirical stretched-exponential (Kohlrausch) fitting. Adapted with permission from Ref.~\cite{Cui_dielectric}, Copyrights American Physical Society (2019).}
\end{figure}

The calculation implemented a model VDOS for a random network of harmonically interacting particles~\cite{Cui_dielectric} since experimental data for the VDOS of the glycerol system were not available. 
The model VDOS however also features a strong boson peak near the glass transition, which modulates the 
$\alpha$ wing asymmetry. 

In later work on orientationally-disordered (plastic) crystals of the fluorinated molecules Freon112 and Freon113, the VDOS measured experimentally via Inelastic Neutron Scattering for the same molecular systems was used in Eq. (29) which led to a significant improvement in the quality of the fitting, results are shown in Fig. (10).

\begin{figure}
\centering{
\includegraphics[height=5.5cm ,width=8cm]{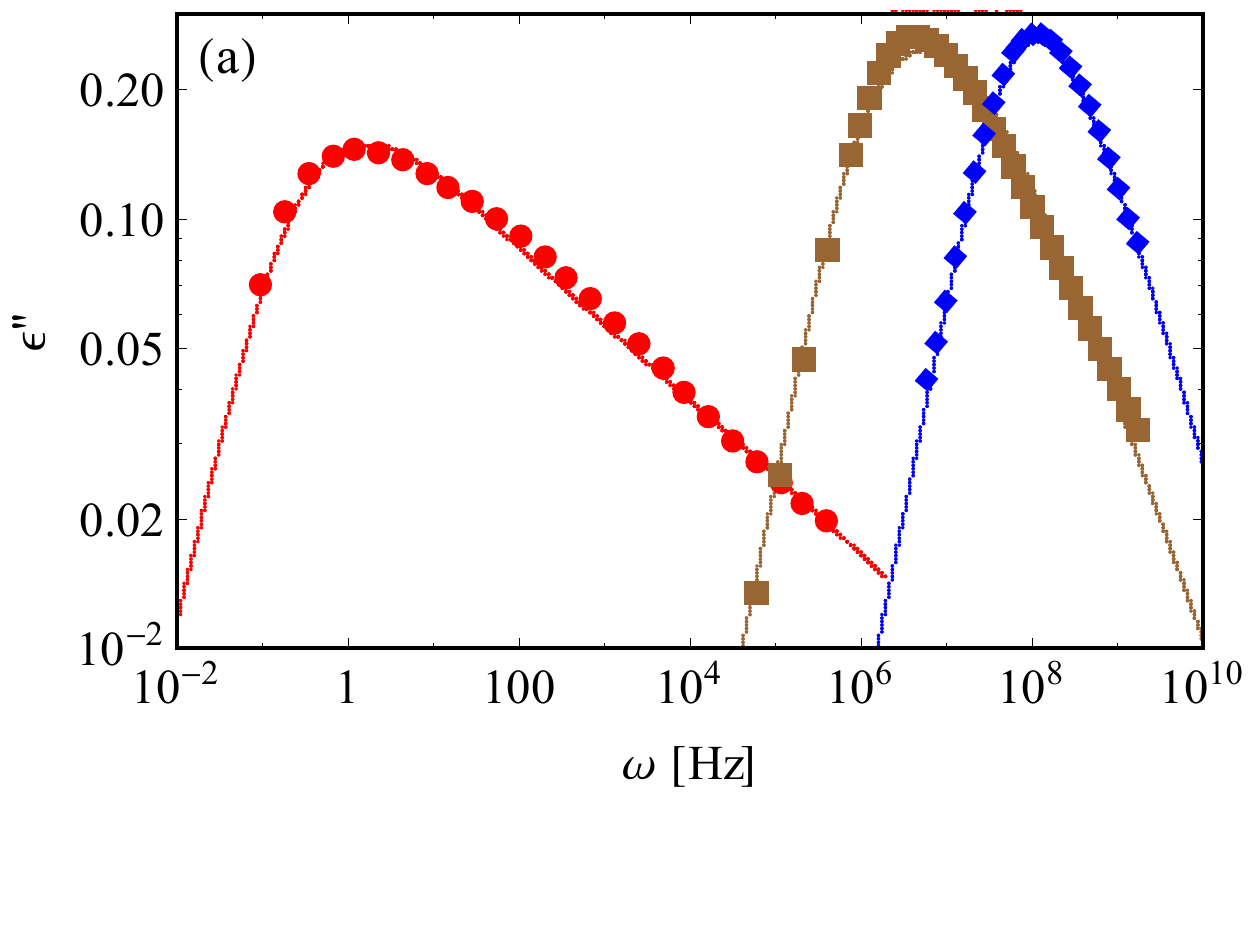}}\hfill
{
\includegraphics[height=5.5cm ,width=8cm]{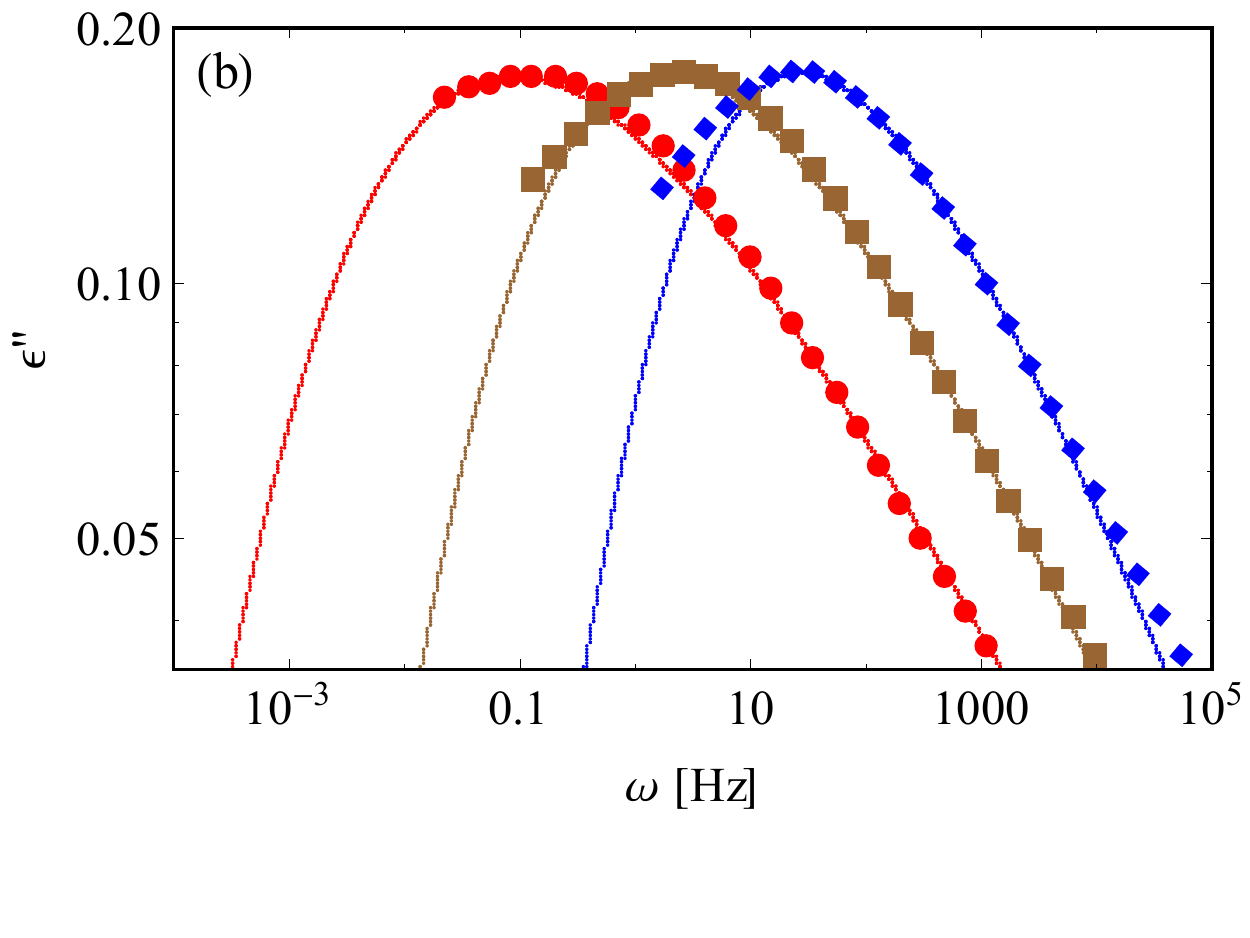}}
\caption{Fitting of experimental data using Eq. (29) (solid lines) for Freon 112 (a) at 91 K (red circles), 115 K  (brown squares) and 131 K (blue diamonds) and for Freon 113 (b) at 72 K (red circles), 74 K (brown squares) and 76 K (blue diamonds). A
rescaling constant was used to adjust the height of the curves since the data
are in arbitrary units. Experimental data for Freon 112 were taken from
Ref.~\cite{Pardo2006}, while data for Freon 113 were taken from
Ref.~\cite{Vispa2017}.}
\label{fig:epsF112&F113}
\end{figure}

Furthermore, the continuous eigefrequency spectrum of dynamical coupling constant $c_\alpha$ in the Hamiltonian Eq. (19) could be extracted from the fitting. This spectrum provides information on the strength of dynamical coupling (anharmonic interactions) in a certain energy window. The analysis gave evidence for a  bump in the energy window $2-5$ THz, which may correspond to secondary $\beta$ relaxation. The bump is significantly larger (by a factor 2) for Freon112 than for Freon113, which  is consistent with the fact that Freon112 has more long-range interactions and anharmonicity than Freon113. This could lead to medium-range dynamical  coupling associated with clustered motions responsible for the secondary $\beta$ relaxation.
Hence, this approach, by allowing the identification of the vibrational energy window where the secondary relaxation takes place, opens up the possibility of isolating the vibrational modes associated with $\beta$ relaxation. This is a new and active line of research that is being tested further on different systems.

\subsection{The link between boson peak and stretched-exponential relaxation}
The above relationship Eq. (29) establishes a direct link between the VDOS and the relaxation through the loss dielectric modulus. 
Through this equation it is therefore possible to study how the boson peak in the VDOS correlates with stretched-exponential relaxation. In order to do this it is best to work with a uniform constant (Markovian) friction $\nu=const$. 

In the time domain, the relationship reads as:
\begin{multline}
\epsilon(t)=B+{}\\
\int_0^{\omega_D}
A\frac{D(\omega_p)}{2K}\left(\frac{e^{(K-\nu/2m)t}}{K-\nu/2m}+\frac{e^{-(K+\nu/2m)t}}{K+\nu/2m}\right) d\omega_p,
\end{multline}
where $K \equiv \sqrt{\frac{\nu^2}{4m^2}-(C\omega_p)^2}$ (please note that there is a difference of a minus sign in the definition of $K$ in Ref.~\cite{Cui_dielectric} due to a typo).
The relationship is plotted in Fig. 11 below in comparison with the dielectric relaxation given by a pure stretched-exponential (Kohlrausch) function. 
It is clear that Eq. (30) is able to reproduce the stretched-exponential relaxation (even without non-Markovianity), as the dielectric relaxation is given by an integral where simple exponentials are weighted by a distribution, the latter being nothing else than the VDOS. One can compare, indeed, Eq. (30) with the generic expression Eq. (1) for the emergence of stretched-exponential relaxation from the superposition of simple exponentials with a suitable distribution of relaxation times (provided by the VDOS as explained below).

\begin{figure}
\includegraphics[height=5.7cm,width=8.0cm]{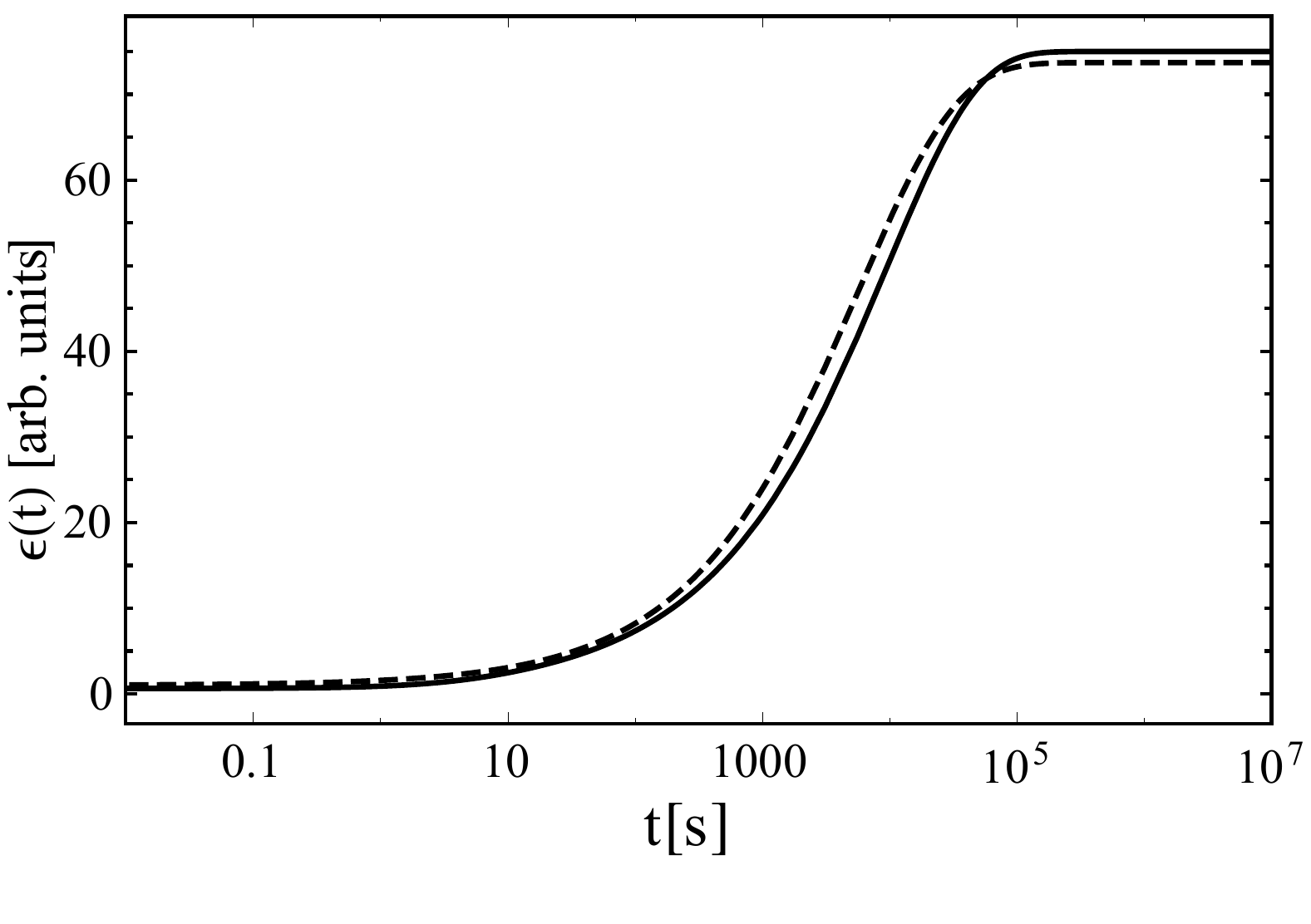}
\caption{Comparison between Eq. (30) (solid line) and a pure stretched-exponential dielectric relaxation in the time domain Eq. (33) (dashed line) with $\beta=0.56$, where the VDOS is the one of a harmonic random network featuring a prominent boson peak~\cite{Milkus}. Adapted with permission from Ref.~\cite{Cui_dielectric}, Copyrights American Physical Society (2019).}
\end{figure}

In Eq. (30), the simple exponentials can give rise to either decaying or increasing exponential behaviour in time, or oscillatory behaviour, depending on the choice of the parameters. For strong damping, decaying exponential behaviour in time dominates the other behaviours, and the leading term can be written in the form $\exp((K-\nu/2m)t)$. Upon Taylor expanding to second order in $1/\nu$, we obtain the following approximate equality (neglecting higher order terms in $1/\nu$):
\begin{equation}
\sqrt{\frac{\nu^2}{4m^2}-\left(C\omega_p \right)^2}-\nu/2m  \approx -C^{2}\frac{m\omega_{p}^{2}}{\nu}
\end{equation}
from which, upon replacing back in the exponential, we obtain
\begin{equation}
e^{(K-\nu/2m)t} \approx e^{-t/\tau}, ~~~ \textrm{with} ~~ \tau=\frac{1}{C^{2}}\frac{\nu}{m\omega_{p}^{2}} 
\end{equation}
which recovers Eq. (16) derived by Tobolsky and co-workers within a constant factor $1/C^{2}$. 

However, due to the distribution of relaxation times/eigenfrequencies provided by $D(\omega_{p})$ in Eq. (30), the resulting behaviour for $\epsilon(t)$ will be stretched exponential:
\begin{equation}
\epsilon(t) \approx B- A' \exp(-t/\tau)^{\beta} 
\end{equation}
as demonstrated in Fig. 11. 

This result establishes an important point:\\
\\
\textit{stretched-exponential relaxation can be recovered within Eq. (1) from simple exponential decay with a distribution of relaxation times that is provided by the VDOS. }\\
\\

In previous work, Eq. (29) evaluated using VDOS  for a model system with variable boson peak has been shown to provide a direct correlation between boson peak height and the extent of $\alpha$  wing asymmetry, in that a larger boson peak translates into a wider or more asymmetric $\alpha$ relaxation wing~\cite{Cui_PLA}. 

Finally, in terms of future work it will be interesting to explore whether a similar mechanism like Eqs. (30)-(33) leading to stretched-exponential relaxation starting from a VDOS with a boson peak, might be active in crystals as well. On one hand, recent work has highlighted, both experimentally~\cite{Moratalla,Jezowski} and theoretically~\cite{Baggioli_PRL},  that a VDOS with a boson peak does arise also in perfect crystals thanks to anharmonic diffusive-like damping. On the other  hand, experimental results were presented in the past which demonstrate stretched-exponential behaviour in the dielectric relaxation of perfect crystals, see the example of $\textrm{TiBO}_{3}$ crystals~\cite{Bokov}. Hence, future work could be directed towards understanding whether a boson peak due anharmonicity in perfect crystals may also lead to stretched-exponential relaxation in e.g. dielectric relaxation. 

\section{A new mystery in the atomic relaxation of metal alloys: compressed exponential relaxation}
One of the most puzzling phenomena that has been observed in experimental measurements on glass-forming metal alloys in recent years is the crossover, right at the $T_g$  from the stretched-exponential relaxation in the supercooled liquid to a compressed exponential relaxation below the $T_g$~\cite{Ruta}, as shown in Fig. 12. The values of the compressed exponent (close to $1.3$) remain constant as a function of temperature below $T_g$, while it varies with the sample aging. The characteristic relaxation time also displays a marked dependence on sample age. 
Above $T_g$, the stretching exponent decreases slightly upon approaching $T_g$ from above, which points towards a narrowing of the distribution of relaxation times, along the interpretation presented above. 
Below $T_g$, however, the compressed exponential behaviour cannot be interpreted within any of the theories or frameworks illustrated above. 

\begin{figure}
\includegraphics[height=5.7cm,width=8.0cm]{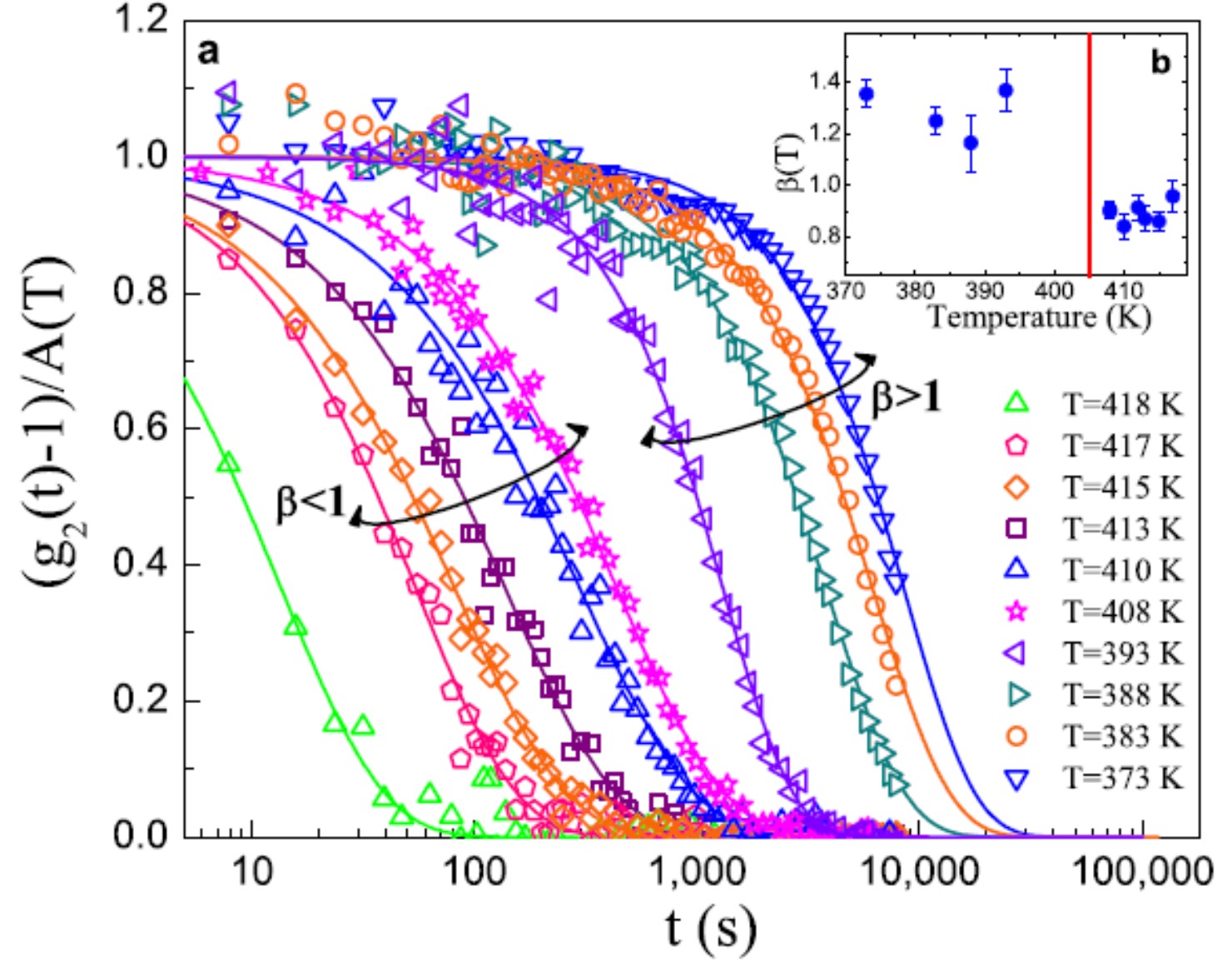}
\caption{Time-dependence of correlation functions measured via X-ray photon correlation spectroscopy on $\textrm{Mg}_{65}\textrm{Cu}_{25}\textrm{Y}_10$ metal alloys, across the glass transition $T_{g}$. The inset reports the values of the stretched/exponent $\beta$ across the transition, which clearly shows the crossover from compressed behaviour ($\beta >1$) below $T_g$ to stretched behaviour ($\beta <1$) above $T_g$. The inset shows the values of exponent $\beta$ across $T_{g}$. Adapted with permission from Ref.~\cite{Ruta}, Copyrights American Physical Society (2019).}
\end{figure}

\subsection{Bouchaud-Pitard elastic dipole relaxation model}
A theoretical explanation of compressed exponential behaviour that has been invoked to interpret these results is based on the relaxation of internal stresses associated with elastic dipoles, within the theory of Pitard and Bouchaud~\cite{Bouchaud}, initially developed for colloidal gels. Compressed exponential relaxation has been observed many times in soft matter, for example in colloidal and jammed systems~\cite{Cipelletti,DelGado,McKenna,Douglas}, normally together with a ballistic relationship between relaxation time and wavevector, $\tau \sim q^{-1}$. 
In this theory, the starting point is the equation of elastodynamics for the displacement field of a (visco)elastic medium in the presence of a finite concentration of point elastic dipoles, and with some viscous damping. The main assumption is that there exists a characteristic time-scale $\theta$ over which a single dipole relaxation event comes to completion. The intensity of the dipole decays on this time-scale. If the time interval which enters the time-correlation functions is much smaller than $\theta$, the solution to the elastic field equations for the displacement yields an intermediate scattering function which behaves like a compressed exponential with $\beta= 3/2$, not far from the value measured experimentally~\cite{Ruta}. 

From a physical point of view, this interpretation is appealing because, right at $T_g$, and this is especially true for metallic glasses where the quench rate is extremely high, internal stresses develop suddenly as the system becomes rigid. This implies that as the structure freezes in, many atoms will find themselves away from the interaction minima with their neighbours, which means that the force obtained as the derivative of the interatomic potential is non-zero, hence there will be a local internal stress associated with every atom displaced from the potential minimum with its neighbour. Each of these interatomic stresses thus behaves like a local elastic dipole, which lends credibility to the elastic dipole model to explain compressed exponential relaxation in metallic glasses. This approach rooted in elasticity could also explain the ballistic (or acoustic-like) dependence of the relaxation time on wavevector, $\tau \sim q^{-1}$. 

It should also be remarked that, as pointed out in a review by Bouchaud~\cite{Bouchaud_review}, superdiffusive models which are often invoked to explain ballistic-type compressed exponential relaxation, in fact predict just a simple exponential relaxation in time, as is the case for L\'{e}vy flights.

\subsection{Superposition of collisional Gaussian functions}
A different mechanism by which compressed exponential relaxation may arise could possibly be by superposition of Gaussian functions, as suggested in Ref.~\cite{Hansen}, in a way similar to how a superposition of simple exponentials yields a stretched-exponential as outlined above. In the case of metal alloys, the various Gaussian functions, with different parameters, could correspond to different atomic species or even to rigid atomic clusters. The ballistic collisional relaxation of an atomic species is known to be given by a Gaussian function of time ~\cite{Verlet}, and the presence of multiple atomic species (or even rigid clusters) in metal alloys could lead to a superposition of Gaussians and hence to a compressed exponential according to the mathematical model presented in \cite{Hansen}. Future work using MD simulation data could be addressed to verify this possible explanation, which, for the case of metal alloys, is proposed here for the first time. 

\subsection{Problems to be solved}
In future work, still, several questions remain to be answered. First of all, what is the relationship between compressed exponential relaxation in the intermediate scattering function and the relaxation behaviour observed in mechanical spectroscopy/rheology (i.e. the Fourier transform of $G''(\omega)$)? Experimental studies indicate that stretched-exponential (Kohlrausch) behaviour in the mechanics persists also below $T_g$~\cite{Evenson_Ruta_2017}, which  would then call for additional modelling efforts in order to explain how compressed exponential behaviour in the intermediate scattering function could be compatible with Kohlrausch stretched behaviour in $G''$ and in $G(t)$. A possible explanation is suggested in Ref.~\cite{Cui_metal} and Section VI.C.2 of the present review, and in particular by Eq. (25) above, which suggests that the exponent of the memory function is actually twice the exponent in the intermediate scattering function. 
 
An interesting question is how the compressed exponential behaviour connects to the VDOS. In recent work on polymer glasses, it has been shown that instantaneous normal modes (INMs), see Sec. IB and Fig. 1 above, are closely connected with residual or internal stresses~\cite{Voigtmann_residual,Palyulin}. Hence, in future work, it could be interesting to look for signatures of the internal stresses in inelastic scattering data that could be linked to INMs from MD simulations. 

Another important question is related to the aging dependence of both the relaxation time and the compressed exponent~\cite{Evenson_PRL,Ruta_review,Evenson_2017}. Simulation and experimental work in this area should be addressed to verifying the assumptions on waiting times for elastic dipole relaxation within the Bouchaud dipole model, and the latter could be extended to fully account for the aging dependence of the compressed exponential behaviour and for the observed intermittency of atomic-relaxation processes in aging dynamics. 

\section{Summary}
We have reviewed results concerning the structural relaxation processes in supercooled liquids and glasses (with particular attention to metal alloys) and the relation thereof with vibrational properties. In spite of the fact that the two fields have developed almost independently of one another for a long time, several links between relaxation and vibrational behaviour have emerged in recent work. These could be summarized schematically as follows:

(i) Different approaches (Fluctuating Elasticity, Mode-Coupling theory, and lately the Nonaffine Generalized Langevin Equation framework) concur in indicating dynamical heterogeneity as the root cause of stretched exponential relaxation behaviour. Dynamical heterogeneity provides a spectrum of relaxation processes, from the superposition of which the stretched exponential behaviour arises. A further connection between vibrational properties and structural relaxation is provided by the linewidth of vibrational excitations measured with various techniques and its frequency dependence that involves a Maxwell viscoelastic crossover with a single relaxation time in the hydrodynamic limit~\cite{Baldi_anharmonic,Giugni}.

(ii) Growing consensus in recent theoretical work from a number of different authors is being gathered which indicates that the vibrational density of states (VDOS) of amorphous solids is characterized by a crossover from ballistic phonons at large wavelength into a regime of quasi-localized excitations, due to disorder-induced scattering becoming important at a length scale at which the phonon wavelength is comparable with the mean free path~\cite{Schirmacher_SciRep}. This crossover gives rise to a deviation from the Debye law and gives rise to the boson peak. Recent work~\cite{Baggioli_random} has revealed a deeper link between the quasi-localized excitations and random matrix statistics of the eigenvalues of the Hessian. In turn, the random matrix statistics can provide a natural explanation for the linear-in-$T$ anomaly in the specific heat of glasses at low temperature, as shown in Ref.~\cite{Baggioli_PRR}.
Finally, the boson peak has been observed also in highly-ordered crystals in recent experimental work on halomethanes~\cite{Moratalla}, and one cannot rule out similar results in metals. The mechanism in this case~\cite{Baggioli_PRL} is controlled by a similar Ioffe-Regel crossover from ballistic phonons to quasi-localized excitations, with the difference that localization is induced by anharmonicity-induced scattering (present also in perfect crystals), as opposed to "harmonic" disorder-induced scattering active in glasses. 

(iii) The NALD-GLE approach has directly identified the spectrum of relaxation processes, which controls stretched-exponential behaviour in mechanical and dielectric relaxation, with the VDOS, via Eq. (24) and Eq. (29) above. A similar connection, although less direct, emerges from the FE framework where the same Green's function with quenched disorder in the elastic modulus controls both the boson peak in the VDOS and the Kohlrausch stretched-exponential behaviour in mechanical relaxation;

(iv) An old mystery in the VDOS of metal alloys, i.e. the observation of a $4/3$ power-law regime in the VDOS in an energy window close to the boson peak~\cite{Suck1980}, could be explained (as suggested in this review for the first time, to our knowledge) by means of the Alexander-Orbach fracton model of vibrational excitations on fractal structures, by identifying the latter with the fractal network of activating units (mobile atoms undergoing medium-range jumps), recently observed in metal alloys~\cite{Ma2008,Yang2017}; in future work, such power-laws could be further put in relation with power-laws observed in avalanches and other intermittent phenomena in strained alloys~\cite{Lagogianni};

(v) Secondary or $\beta$ relaxation in metal alloys has been shown in recent work to be associated with clusters of mobile atoms which undergo coordinated motions~\cite{Samwer_review}. The visibility and extent of the $\beta$ peak may be related to atomic size mismatch in the alloy, with the high-mobility clusters being composed to a larger extent by the smallest/lightest atomic species~\cite{CuiPRB2018}. Future work could be addressed to identifying which specific vibrational modes are associated with atomic motions involved in $\beta$ relaxation~\cite{CuiPRE2018}, using the relaxation time-eigenfrequency equivalence relations presented in this review. This identification may bring new insights into the nature of $\beta$ relaxation at the microscopic level and new routes for its mathematical modelling.

(vi) The recent observation of compressed exponential relaxation in the intermediate scattering function of metal alloys below $T_g$ and the crossover from compressed to stretched relaxation upon crossing $T_g$~\cite{Ruta} have triggered new fundamental questions. While compressed exponential behaviour could be explained with the Pitard-Bouchaud model of relaxing elastic dipoles connected with internal stresses, it remains to be understood why the mechanical relaxation remains of stretched exponential type below $T_g$, and what the connection would then be between stretched exponential relaxation in the mechanics and compressed exponential relaxation in the intermediate scattering function. Solving this crucial issue may call for a rethinking and revisiting of the Pitard-Bouchaud model, its extension to link with the dynamics of intermittent events, clustering of atomic motion, and its combination with models of nonaffine deformation of metal alloys~\cite{Cui_metal}. 

Important topics that have been left out in this review, but which are also important in the current debate are the following: (i) the relationship between $\alpha$ and $\beta$ relaxation, in particular as a function of temperature, with the secondary peak which moves farther apart from the primary peak upon lowering the temperature, a topic that has been investigated with the Ngai coupling model~\cite{Ngai} for the coupling between the two characteristic relaxation times; (ii) the complex relationship between ageing, relaxation and vibrational properties which is hitherto completely unexplored from a theoretical point of view, although a large body of experimental data on ageing in metal alloys is available~\cite{Ruta_review,Wang_review2019}; (iii) the physics of relaxation in the nonlinear regime, i.e. under large amplitudes of oscillatory strain (in metal alloys) or electric field (in organic glass formers), which is a topic of great interest for current research~\cite{Richert}.

\begin{acknowledgements}
Many useful discussions with W. Goetze, G. Ruocco, G. Baldi, K. Samwer, J.-L. Tamarit, Th. Voigtmann, E. Pineda, K. Trachenko, Z. Evenson, and with E. M. Terentjev are gratefully acknowledged. This work was supported by contract W911NF-19-2-0055 with US Army Research Office (ARO) and Army Research Laboratory (ARL).
\end{acknowledgements}

\bibliographystyle{apsrev4-1}
\bibliography{relaxation-vibration}

\end{document}